\newcommand{\aeff}{\Omega_\textrm{e}}

\newcommand{\mkm}{\mathrm{km}}
\newcommand{\m}{\mathrm}
\newcommand{\degree}{\ensuremath{^\circ}}
\newcommand{\Db}{D_\mathrm{b}}

\documentclass{emulateapj}
\usepackage{graphicx}
\usepackage{natbib}
\usepackage{amssymb}

\begin{document}
\title{The TAOS Project: Upper Bounds on the Population of Small KBOs
  and Tests of Models of Formation and Evolution of the Outer Solar
  System}

\author{
F.~B.~Bianco\altaffilmark{1, 2, 3, 4},
Z.-W.~Zhang\altaffilmark{5,6},
M.~J.~Lehner\altaffilmark{5,3,4},
S.~Mondal\altaffilmark{7},
S.-K.~King\altaffilmark{5},
J. Giammarco\altaffilmark{8,9},
M. J. Holman\altaffilmark{4},
N.~K.~Coehlo\altaffilmark{10},
J.-H.~Wang\altaffilmark{5,6},
C.~Alcock\altaffilmark{4},
T.~Axelrod\altaffilmark{11},
Y.-I.~Byun\altaffilmark{12},
W.~P.~Chen\altaffilmark{6},
K.~H.~Cook\altaffilmark{13},
R.~Dave\altaffilmark{14},
I.~de~Pater\altaffilmark{15},
D.-W.~Kim\altaffilmark{12,14},
T.~Lee\altaffilmark{5},
H.-C.~Lin\altaffilmark{5},
J.~J.~Lissauer\altaffilmark{16},
S.~L.~Marshall\altaffilmark{17},
P.~Protopapas\altaffilmark{14,4},
J.~A.~Rice\altaffilmark{10},
M.~E.~Schwamb\altaffilmark{18},
S.-Y.~Wang\altaffilmark{5} and
C.-Y.~Wen\altaffilmark{5}
}
\altaffiltext{1}{Department of Physics, University of California Santa Barbara, 
Mail Code 9530,  Santa Barbara CA 93106-9530 }\email{fbianco@lcogt.net}
\altaffiltext{2}{Las Cumbres Observatory Global Telescope Network, Inc.
6740 Cortona Dr. Suite 102, Santa Barbara, CA 93117}
\altaffiltext{3}{Department of Physics and Astronomy, University of Pennsylvania, 209 South 33rd Street, Philadelphia, PA 19104}
\altaffiltext{4}{Harvard-Smithsonian Center for Astrophysics, 60 Garden Street, Cambridge, MA 02138}
\altaffiltext{5}{Institute of Astronomy and Astrophysics, Academia Sinica. P.O. Box 23-141, Taipei 10617, Taiwan}
\altaffiltext{6}{Institute of Astronomy, National Central University, 300 Jhongda Rd, Jhongli 32054, Taiwan}
\altaffiltext{7}{Aryabhatta Research Institute of Observational Sciences (ARIES), Manora Peak, Nainital-263 129, INDIA}
\altaffiltext{8}{Department of Astronomy and Physics,
 Eastern University 1300 Eagle Road Saint Davids, PA 19087}
\altaffiltext{9}{Department of Physics, Villanova University, 800 Lancaster Avenue, Villanova, PA 19085}
\altaffiltext{10}{Department of Statistics, University of California Berkeley, 367 Evans Hall, Berkeley, CA 94720}
\altaffiltext{11}{Steward Observatory, 933 North Cherry Avenue, Room N204 Tucson AZ 85721}
\altaffiltext{12}{Department of Astronomy, Yonsei University, 134 Shinchon, Seoul 120-749, Korea}
\altaffiltext{13}{Institute for Geophysics and Planetary Physics, Lawrence Livermore National Laboratory, Livermore, CA 94550}
\altaffiltext{14}{Initiative in Innovative Computing at Harvard, 60 Oxford St., Cambridge MA 02138}
\altaffiltext{15}{Department of Astronomy, University of California Berkeley, 601 Campbell Hall, Berkeley CA 94720}
\altaffiltext{16}{Space Science and Astrobiology Division 245-3,  NASA Ames Research Center, Moffett Field, CA, 94035}
\altaffiltext{17}{Kavli Institute for Particle Astrophysics and Cosmology, 2575 Sand Hill Road, MS 29, Menlo Park, CA 94025}
\altaffiltext{18}{Division of Geological and Planetary Sciences, California Institute of Technology, 1201 E. California Blvd., Pasadena, CA
 91125}

\begin{abstract}
  We have analyzed the first 3.75 years of data from TAOS, the
  Taiwanese American Occultation Survey. TAOS monitors bright stars to
  search for occultations by Kuiper Belt Objects (KBOs). This dataset
  comprises $5\times10^5$ star-hours of multi-telescope photometric
  data taken at 4 or 5~Hz. No events consistent with KBO occultations
  were found in this dataset. We compute the number of events expected
  for the Kuiper Belt formation and evolution models of
  \citet{2005Icar..173..342P}, \citet{2004AJ....128.1916K},
  \citet{2009P&SS...57..201B}, and \citet{fraser09}. A comparison with
  the upper limits we derive from our data constrains the parameter
  space of these models. This is the first detailed comparison of models
  of the KBO size distribution with data from an occultation
  survey. Our results suggest that the KBO population is comprised of
  objects with low internal strength and that planetary migration
  played a role in the shaping of the size distribution.
\end{abstract}


\section{Introduction}\label{sec:intro}
\setcounter{footnote}{0} 

The Kuiper Belt has been shaped by accretion and disruption processes
throughout the history of the Solar System.  With small orbital
eccentricities, the relative velocities of the objects in the early
Kuiper Belt were sufficiently low to allow accretion processes to form
kilometer and much larger objects. Later, the velocity dispersion
increased, possibly as the KBO population was stirred up by the
gravitational effects of the larger planets and planetoids. Only large
objects were then able to continue growing through impacts, whereas
collisions among smaller bodies resulted in disruption. The details of
these processes depend on the internal strength of the KBOs and on the
orbital and dynamical evolution of the gas giant planets.  The size
distribution of KBOs, therefore, contains information on the internal
structure and composition of the KBOs -- and hence information on the
location and epoch in which they formed -- and on planetary
migration \citep[and references therein]{2008ssbn.book..293K}.
Direct observations have detected KBOs as faint as magnitude
$R\sim28.2$ \citep{2004AJ....128.1364B}, which corresponds to a
diameter of about 27~km assuming a 4\% albedo. The large
end side of the KBO size distribution can therefore be characterized
through its brightness distribution. The latter is well described by a
power law $\Sigma(<R)=10^{\alpha (R-R_0)}$, with an index $\alpha=0.6$
and $R_0=23$ (\citealt{2009AJ....137...72F}
\citealt{2008arXiv0804.3392F}) for objects brighter than about
$R~=~25$, or $D\sim100~\mkm$.  This is the region of the size spectrum
which reflects the early history of agglomeration.
\citet{2001ApJ...547L..69K} pointed out that the intensity of the
infrared Zodiacal Background sets limits on the extrapolation of a
straight power law to smaller sizes. The relatively shallow size
distribution of Jupiter Family Comets (JFCs,
\citealt{2006Icar..182..527T}), which are believed to originate in the
Kuiper Belt, and the cratering of Triton observed by \emph{Voyager
  2}~\citep{1996AJ....112.1203S}, all point to a flatter distribution
for small KBOs\footnote{The relationship between the cratering of
  Triton and the Kuiper Belt size distribution is questioned by
  \citet{2007Icar..192..135S}.}.  In 2004 evidence surfaced that a
break in the power law occurs at a diameter larger than $10 ~\mkm$:
\cite{2004AJ....128.1364B} conducted deep Hubble Space Telescope
observations with the Advanced Camera for Surveys which led to the
discovery of only 3 new objects fainter than R~=~26, about 4\% of the
number expected from a single power law distribution extrapolated to
$10~\mkm$.  While this work remains the state of the art for deep
direct surveys of the Outer Solar System, recent campaigns have
observed many more faint objects down to magnitude $R=27$, which with
the assumption of a $4\%$ albedo corresponds to about $40~\mkm$ in
diameter\footnote{The magnitude of KBOs is converted into diameter by
  assuming a nominal 4\% albedo throughout the paper, note however
  that \citealt{2008Icar..198..452F} and \citealt{2009AJ....137...72F}
  assumed an albedo of 6\% in their work.}
(\citealt{2008Icar..198..452F}, \citealt{2008arXiv0804.3392F},
\citealt{2008arXiv0809.4166F}, and \citealt{2009AJ....137...72F}).
These recent data allowed them to locate a break in the power law size
distribution at diameters $30\lesssim D \lesssim 120~\mkm$.

The region of the size spectrum between tens of kilometers and meters
in diameter is particularly interesting as models predict here the
occurrence of transitions between different regimes where the binding
energy of KBOs is dominated either by gravity or internal
strength. These transitions would leave a signature in the size
distribution (\citealt{2005Icar..173..342P},
\citealt{2004AJ....128.1916K}, \citealt{2009P&SS...57..201B}, and
references therein).  Occultation surveys allow us to reach farther
then the current limits of direct observations, and into this region
of interest. These surveys monitor background stars in order to detect
the chance alignment of a KBO with a target star, which would generate
a variation in the observed flux of the star.  At distances in the
Outer Solar System (tens to thousands of AU) the signature left in a
lightcurve by the transits of $D\sim 1~\mkm$ objects is dominated by
diffraction.  This technique requires high frequency photometric time
series as the time scale for an occultation by an Outer Solar System
Object is a fraction of a second (\citealt{2000Icar..147..530R},
\citealt{2007AJ....134.1596N}, \citealt{2009arXiv0902.3457B}).  A few
such surveys have been attempted in the past several years and have
recently started reporting results: e.g., \citet{2006AJ....132..819R},
\citet{2007MNRAS.378.1287C}, \citet{2008AJ....135.1039B},
\citet{2008MNRAS.388L..44L}, \citet{2008ApJ...685L.157Z}
--~hereinafter Z08~--, \citet{mmt}, and \citet{ps}.  None of these surveys have
claimed detections in the Kuiper Belt; upper limits have thus been
placed on the number density of KBOs in the sky. 

\citet{2008AJ....135.1039B} set an upper limit to the sky density of
KBOs of $\Sigma_N(D\geq1~\mkm)~\leq~2.8\times10^9~\m{deg}^{-2}$ using
the 40 Hz data from their own survey as well as the 45~Hz data from
\citet{2006AJ....132..819R} and the X-ray data from
\citet{2007MNRAS.378.1287C}.  \citet{mmt} carried out a 30~Hz survey
with Megacam at the MMT setting a more stringent limit of
$\Sigma_N(D\geq1~\mkm)\leq 2.0\times10^8~\m{deg}^{-2}$ and a limit of
$\Sigma_N(D\geq0.7~\mkm)\leq 4.8\times10^8~\m{deg}^{-2}$.  \citet{ps}
reported preliminary analysis of videomode engineering data taken with
the Pan-STARRS system.  

Recently \citet{fgs} reported the detection of a candidate occultation
event consistent with a $D\sim1~\mkm$ KBO in the analysis of archival
guiding data from HST, and an estimate of the sky density of KBOs of
$\Sigma_N(D\geq0.5~\mkm) ~=~2.1^{+4.8}_{-1.2}\times10^7~\m{deg}^{-2}$.

\begin{deluxetable}{lll}
\tablecolumns{3}
\tablewidth{0pc}
\tablecaption{Dataset parameters (3--telescope data)}
\tablehead{   & Z08  & this work  }
\startdata
Start Date & 2005 February 7 & 2005 February 7\\ 
End Date & 2006 December 31 & 2008 August 2\\
Light-curve sets & 110,554 & 366,083 \\
Exposure (star--hours) & 152,787 & 500,339\\
Triplets\tablenotemark{a}& $ 2.6 \times 10^{9}$ & $ 9.0\times 10^{9}$
\enddata
\tablenotetext{a} {Multi--telescope measurements.}
\label{tab:dataset}
\end{deluxetable}
The Taiwanese American Occultation Survey (TAOS) has been operating
since 2005 with two, three, and now four telescopes simultaneously
taking stellar photometry at 5~Hz\footnote{A small subset of early
  data was collected at 4~Hz cadence, comprising about 5\% of the data
  analyzed in this work.}.  The analysis of the first two years of
TAOS reported no detections (Z08) and an upper limit was derived to
the slope of the small size end of the size spectrum.  The TAOS system
is described in detail in \citet{2008arXiv0802.0303L}.  Using $50~
\mathrm{cm}$ aperture robotic telescopes in simultaneous observations
and observing with the relatively low cadence compared to the
aforementioned occultation surveys, TAOS was designed to address the
km-size region of the KBO size spectrum. We will show here that the
marginal sensitivity to sub-km objects is more than compensated by the
very large exposure of our star targets. Here we consider the first
3.75 years of TAOS data, a significantly larger dataset than the one
explored in Z08. With these data we are able to constrain Kuiper Belt
formation and evolution models.

In Section \ref{sec:data} we describe the new dataset. In Section
\ref{sec:simul} we briefly describe our detection algorithms, as well
as our efficiency analysis. We also discuss our recovery efficiency
and discuss the most productive strategies for TAOS and the other
occultation surveys, and, in Section~\ref{sec:omega}, we derive the
effective coverage of our survey.  In Section~\ref{otherocc} we derive
model--independent limits to the number of objects in the Kuiper Belt,
and we compare our results with those of similar surveys. In
Section~\ref{sec:models} we briefly describe models for the formation
and evolution of the Kuiper Belts and we then derive and discuss
constraints to these models.  In Section~\ref{sec:JFC} we compare our
upper limits to the estimates on the number of KBOs set by dynamical
simulations for JFC progenitor populations. Finally we summarize and
discuss our findings in Section~\ref{sec:conc}.

\section{3.75 Years of TAOS Data}\label{sec:data}

\begin{figure*}[ht!]
\centerline{\includegraphics[width=0.5\textwidth]{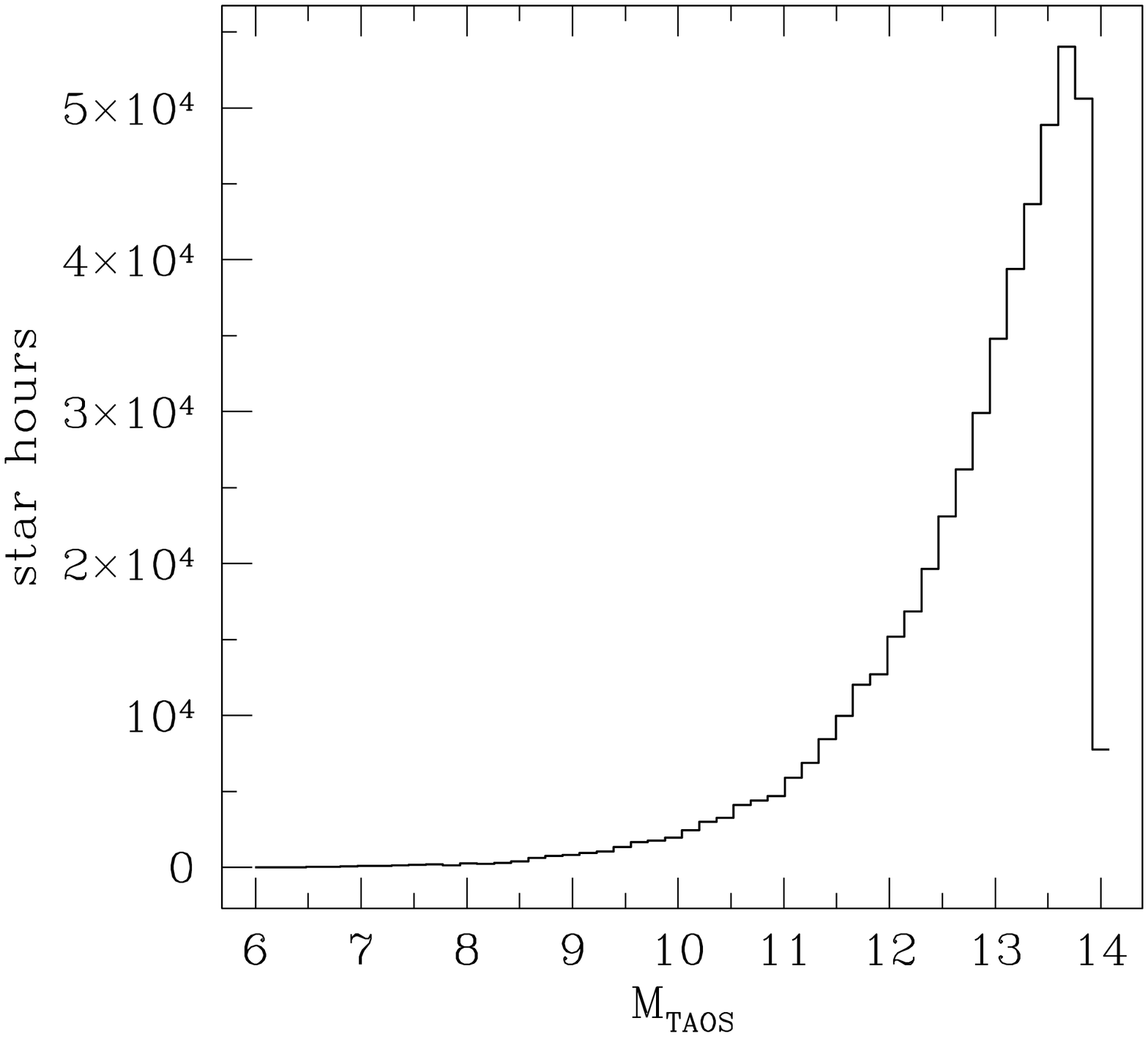}\includegraphics[width=0.5\textwidth]{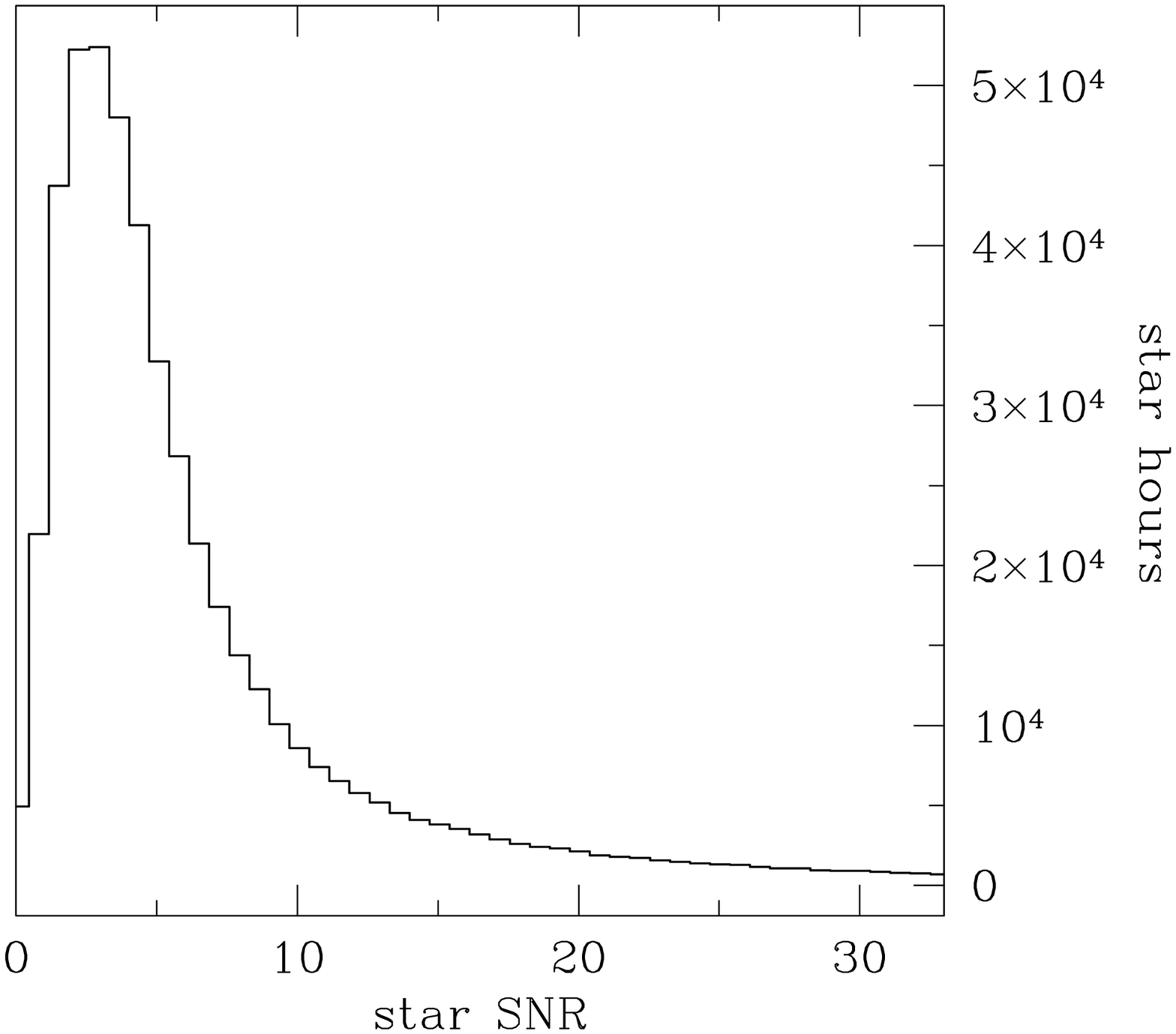}}
\caption{Distribution of magnitudes for the TAOS target stars, (bin
  size 0.16 mag, left).  SNR for the TAOS target stars, averaged over the
  duration of a run and over the three telescopes (bin size 0.73,
  right). A few targets at greater SNR and brighter magnitude,
  amounting to $<5\%$ of the data, are not shown.}\label{fig:r_snr}
\end{figure*}
\begin{figure}[b!]
\centerline{\includegraphics[width=0.5\textwidth]{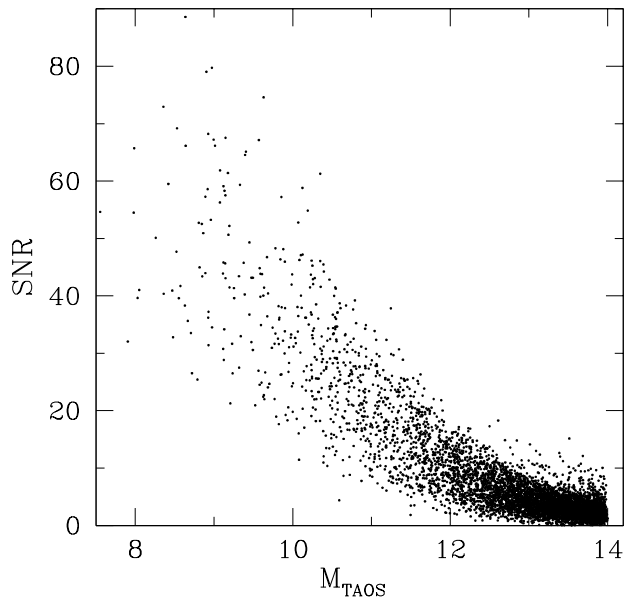}}
\caption{SNR versus TAOS instrumental magnitude $M_\m{TAOS}$; only a random
  sample of 1\% of all stars is shown for clarity.}\label{fig:rvssnr}
\end{figure}

TAOS is a dedicated survey that observes at a cadence of 5~Hz.  The
primary scientific goal of the survey is to estimate or set
constraints on the number of KBOs in the region of the size spectrum
that is currently too small to be observed directly: $D\lesssim
10~\mkm$.

Here we present an expanded analysis of three-telescope TAOS
data\footnote{The fourth telescope, TAOS C, became operational in
  August 2008. The results presented in this paper are based on
  analysis of all of the three-telescope data collected to this
  point.}. These data consist of photometric measurements of target
star fields collected synchronously with all three telescopes.  The
dataset analyzed here was collected between January 2005 and August
2008.  In a previous analysis of a subset of these data, Z08 reported
an upper limit to the size distribution of KBOs under the assumption
of a single power law for small KBOs.  If one models the size
distribution for objects smaller than $D=28~\mkm$, the smallest direct
observation \citep{2004AJ....128.1364B}, as a single power law
${dN}/{dD}~ \propto~D^{-q}$, where $N$ is the surface density of
objects, the slope of the distribution is limited to $q\le4.6$.

Throughout the remainder of this paper, a \emph{data run} refers to a
set of data collected in an uninterrupted observation of any field.
For a single star in the field a set of three lightcurves belonging to
one data run will be referred to as a \emph{lightcurve set}, and each
three-telescope measurement, at a single time point, will be referred
to as a \emph{triplet}. A \emph{star--hour} refers to an hour of
high-cadence, multi-telescope observations on a single target star.

The data set described in this paper amounts to amount to
$5.0\times10^5$ star--hours, while the data set used in Z08
comprises $1.5\times10^5$ star--hours. The details of this dataset, and
of the dataset published in Z08, are summarized in
Table~\ref{tab:dataset}. Over 90\% of our data is collected within
$5\degree$ of the ecliptic plane in order to maximize the rate of
occultations.

\begin{figure*}[t!]                                                                 
\centerline{\includegraphics[width=0.5\textwidth]{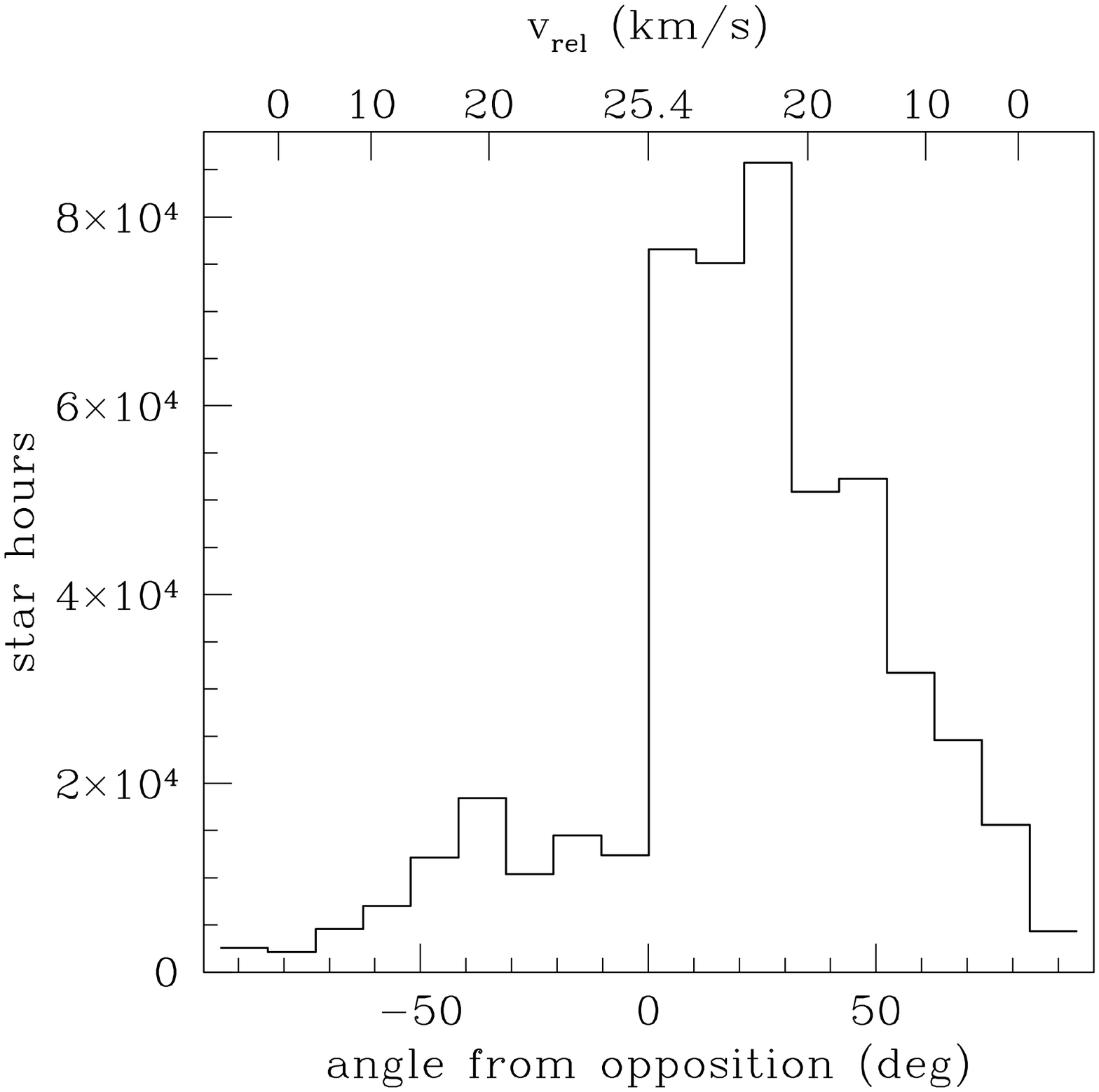}\includegraphics[width=0.5\textwidth]{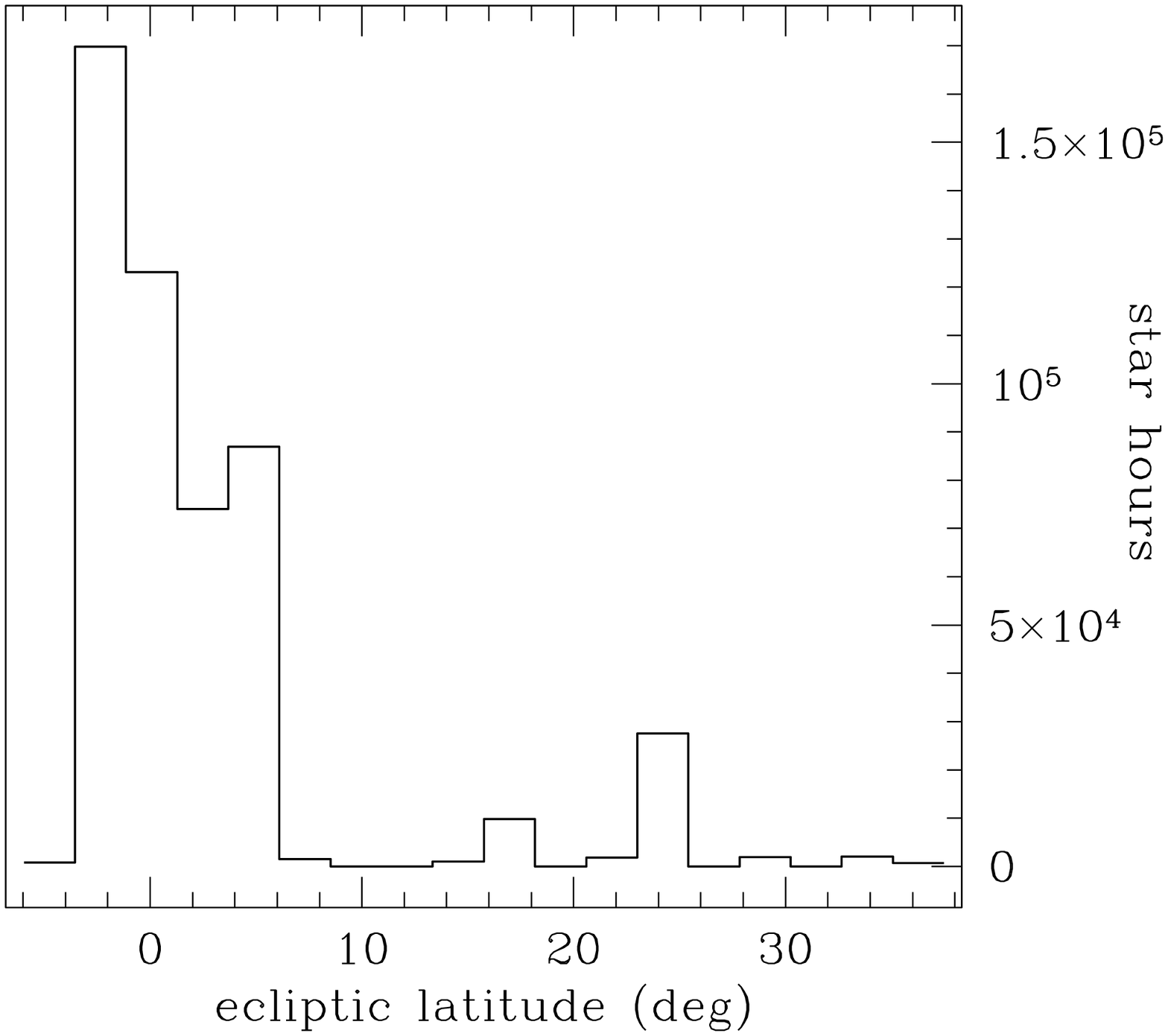}}
\caption{Left: distribution of angles from opposition for the TAOS
  targets. The top axis shows the relative velocity of a KBO at 43 AU,
  given the position of the field. The bin size is 10.5\degree. Right:
  distribution of ecliptic latitude for the TAOS target fields (center
  of the field is assumed), bin size
  $\sim2.5\degree$}\label{fig:oppang}
\end{figure*}

TAOS uses the \emph{zipper--mode} technique to read out the CCD
cameras at high frequency. This method, described in
\cite{2008arXiv0802.0303L}, enables high speed observations across the
3$\Box^\circ$ field of view of the TAOS telescopes, but it
artificially increases the crowding of the field and the background.
In zipper--mode readout each star in the field is represented in a
subsection of the output image -- which we call \emph{rowblock} and
which comprises 76 rows for our 5~Hz data -- so that the field of view
is entirely imaged in each rowblock. Note that the images of different
stars in a rowblock, however, do not necessarily belong to the same
epoch. The zipper--mode readout boosts the sky background by a factor
of 27 at a 5~Hz readout rate. This limits the sensitivity of TAOS to
stars as faint as $M_\m{TAOS} = 13.5$, for which a signal--to--noise
ratio (SNR) of $\sim7$ can be achieved in a dark night.  The magnitude
and SNR distributions for the target stars in our survey are shown in
Figure~\ref{fig:r_snr}. On the left panel, the $x-$axis is the TAOS
instrumental magnitude $M_\m{TAOS}$, which is defined by a regression
on the USNO-B magnitudes to be similar to $R_\m{USNO}$. The
correlation between instrumental magnitude and SNR is shown in
Figure~\ref{fig:rvssnr}. The scatter in the relationship between SNR
and $M_\m{TAOS}$ is due to both changes in the sky background and in
the weather conditions, and to different degrees of crowding in the
fields. In Figure~\ref{fig:oppang}, left, we show the number of
star--hours at different angles from opposition. The top scale
indicates the velocity of a KBO at the center of a field at this
elongation.  We cover a large range of opposition angles; our field
selection algorithm favors ecliptic fields near zenith. Most angles
are positive because the weather at the site tends to improve after
midnight. The right panel of Figure~\ref{fig:oppang} shows the
distribution of ecliptic latitude of our data. The effects of the
angle from opposition on our efficiency and event rate, as well as the
efficiency as a function of magnitude and crowding are discussed
further in Section~\ref{sec:analysis}.

\section{Analysis}\label{sec:simul}

The first step in the analysis is the photometric reduction of the
zipper mode images in the data set.  A custom aperture photometry
package~\citep{TAOS_photpaper} is used to measure the brightness of
each star at each epoch, and the resulting series of flux measurements
are then assembled into lightcurves for subsequent analysis, which is
described in the following subsections.  In Section~\ref{sec:fp} we
describe our detection algorithm and the rejection of false positives.
We then describe the efficiency tests: in Section~\ref{sec:angs} we
describe in detail how we identify the angular size of our target
stars to simulate occultations correctly. In
Section~\ref{sec:analysis} we show how we simulate and implant
occultation events in our lightcurves, and test the behavior of our
efficiency as a function of various parameters relative to the
occultations and to the observing strategy.  Finally, we can derive the
effective coverage of our survey (Section~\ref{sec:omega}).

\subsection{Event Detection and False Positive Rejection}\label{sec:fp} 
The \emph{Fresnel scale} is defined as $F = (\lambda \Delta/2)^{1
  \over2}$ where $\lambda$ is the wavelength of observation, and
$\Delta$ the distance to the occulter~\citep{1987AJ.....93.1549R,
  1980poet.book.....B}.  For optical observations at the distance of
the Kuiper Belt (about 43~AU) the Fresnel scale is $F\approx
~1.4~\mkm$.  Occultation events will therefore exhibit significant
diffraction effects. Occultations are manifested in the lightcurve of
an observed star as an alternation of bright and dark features,
typically with an overall suppression of the flux. Theoretical
occultation lightcurves are shown in
Figure~\ref{fig:sim}.  The signature of an
occultation by a KBO of sub-kilometer size has a duration of about 0.2
second at opposition, and about a second near quadrature.  A typical
KBO occultation is then expected to result in the suppression of the
flux for one or a few consecutive points in a TAOS lightcurve. In
order to ascertain the extra-terrestrial origin of a dip in a
lightcurve TAOS observes simultaneously with multiple telescopes. This
allows us to rule out, on the basis of simple parallax considerations,
atmospheric scintillation phenomena which might mimic an occultation
event and which could be a source of false positives in occultation
surveys, as well as any non-atmospheric phenomena such as birds,
airplanes, etc.  

In order to detect occultations we need to identify brief flux changes
in a star simultaneously observed by all telescopes.  
The statistical significance of a simultaneous low point in our
lightcurves can be assessed rigorously, and the probability of a low
measurement being drawn out of pure noise decreases with the number of
telescopes observing the target, provided that the measurements for
the telescopes are independent. The lightcurves are high-pass filtered
to remove trends due to weather patterns and changes in atmospheric
transparency. High-pass filtering the lightcurves preserves the
information on time-scales relevant to occultation phenomena (one or a
few points in a time series). The implementation of the filter is
described in Z08. The filter produces a time series in which $h(t)$,
the measurement taken at time $t$, represents the deviation from the
local mean of the lightcurve in units of local standard deviation. 
\begin{figure*}[ht!]
\centerline{\includegraphics[width=0.4\textwidth]{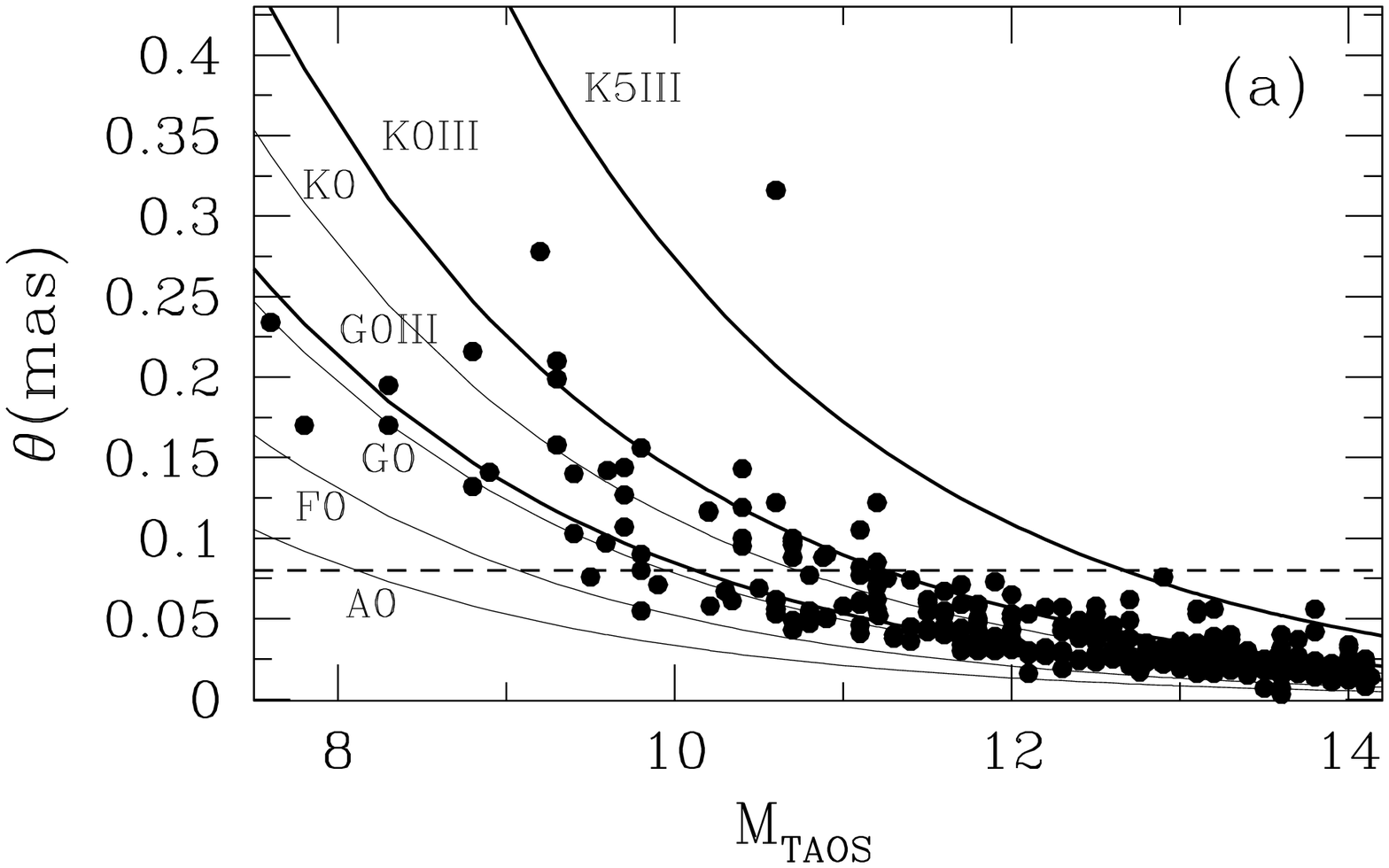}\includegraphics[width=0.4\textwidth]{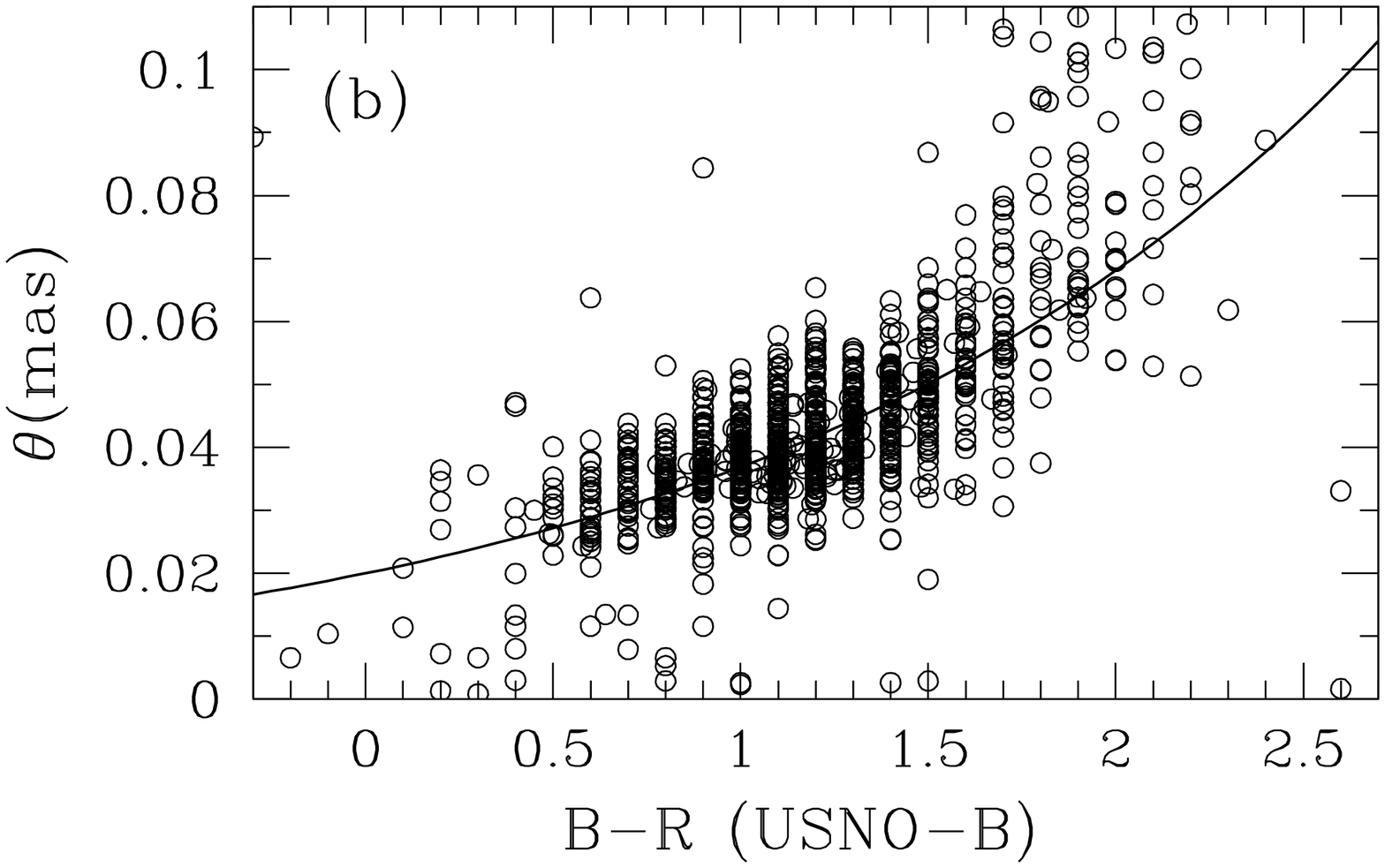}}
\centerline{\includegraphics[width=0.4\textwidth]{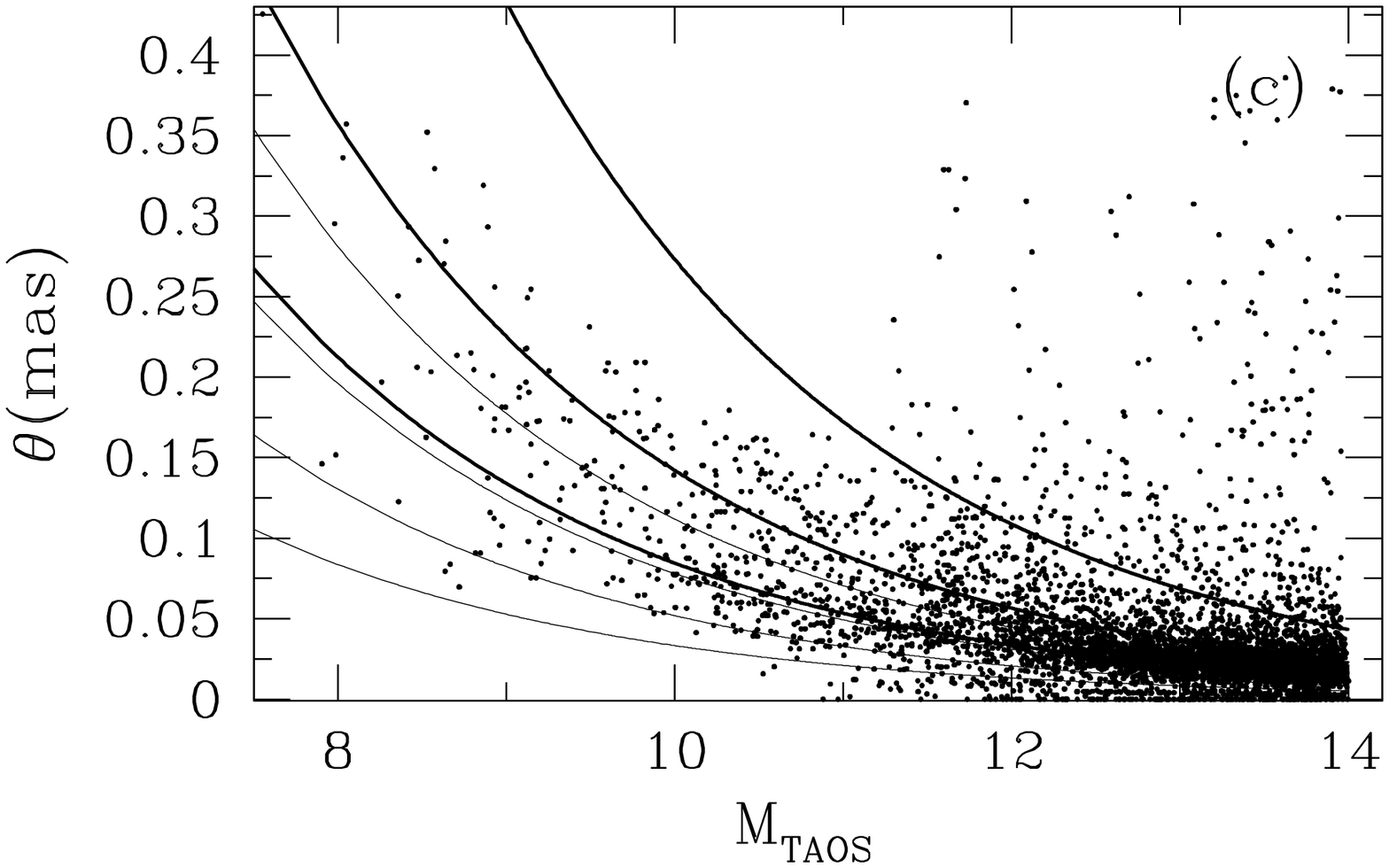}\includegraphics[width=0.4\textwidth]{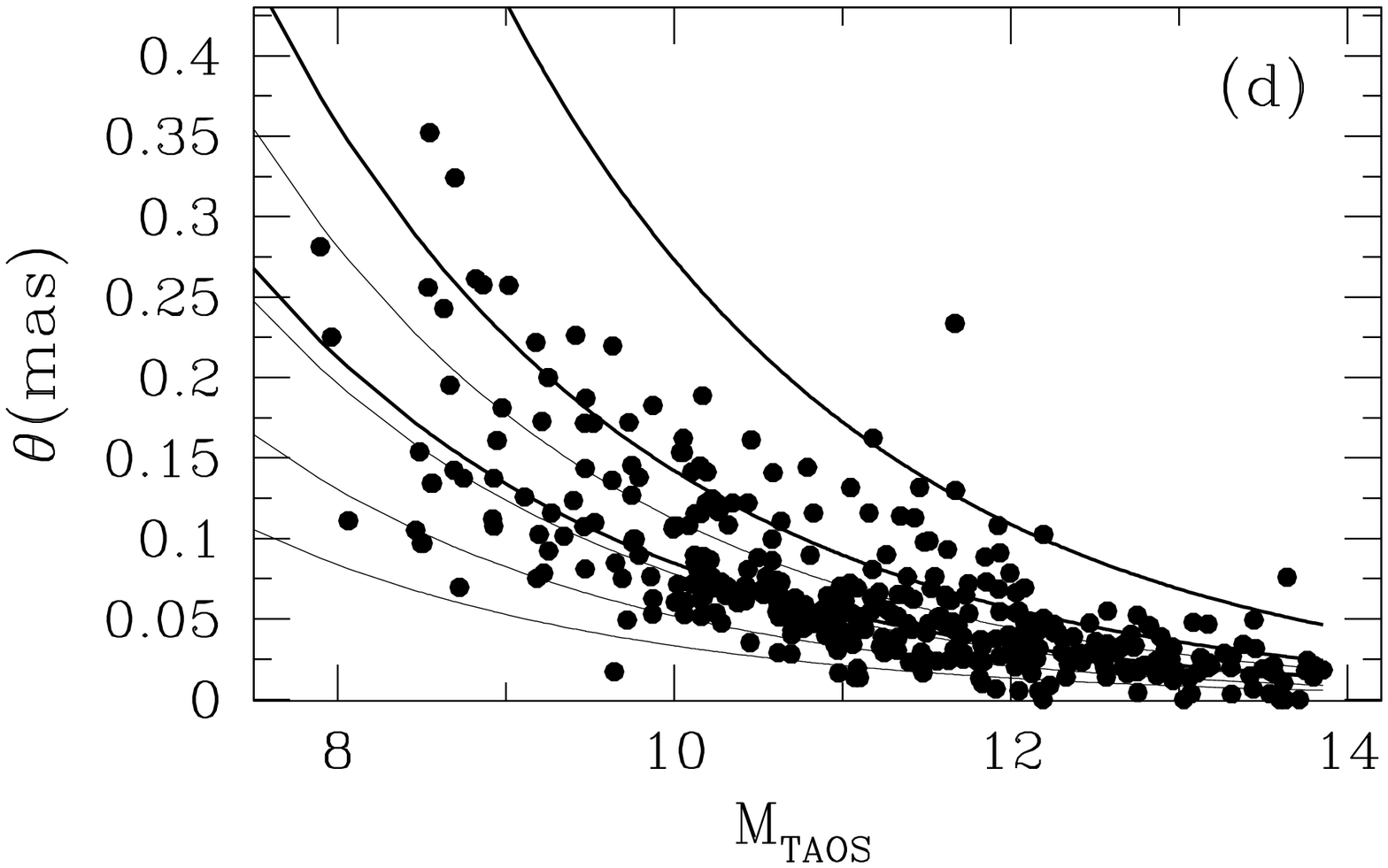}}
\caption{Angular size distribution for a typical TAOS field (field
  120, RA: $13\degree .7$, Dec: $-10\degree.7$) derived from the 2MASS
  $K-J$ colors (a). The curves show the
  theoretical behavior of the angular size for A0, F0, G0, K0 dwarf
  stars (thin lines) and G0, K0, K5 giants (thick lines). The size of
  the Fresnel scale at 43~AU is shown at 0.08~mas (dashed line). Best
  fit to the angular size distribution (b):
  \emph{x}-axis is the USNO-B $B-R$ color and \emph{y} is the angular size
  derived using Equations~\ref{eq:sb}~\&~\ref{eq:angsoumen},
  converting all stars to apparent magnitude $R~=~12$.  Implanted
  angular sizes, derived from USNO-B the $B-R$ color for our
  simulation (c) and angular size of the star for
  which events are recovered (d); a random subset of
  1\% of our data is plotted in the bottom panels.}\label{fig:angsz}
\end{figure*}

To detect events we rank--order the photometric measurement in each of
our lightcurves, from the lowest to the highest flux, independently
for each telescope (labeled A, B and D). The $i$-th point in a
lightcurve will be associated to rank $r_i^\m{T}$ for telescope~T. We
then consider the rank triplets
($r_i^\m{A},~r_i^\m{B},~r_i^\m{D}$). The probability distribution of
the quantity $z_i~=~-\ln\{r_i^\m{A}~r_i^\m{B}~r_i^\m{D}/N_\m{p}^3\}$,
with $N_\m{p}$ the number of points in the lightcurve set, can be
determined combinatorially.  Knowing this, under the null hypothesis
that there is no event in the triplet $i$, we can compute the
probability for a random variable $Z$ arising from this distribution
$P(Z>z_i)~=~\xi$. We set a threshold such that we expect fewer than
0.27 events in our dataset that are due to random fluctuations. For
the dataset discussed in this paper we accept as events all data
points that produce a rank product less likely than $\xi =
3.0\times10^{-11}$ to be drawn from a random distribution.  Note that
events generated by large KBOs, or for observations near quadrature,
would affect more than one point in the lightcurve
(Figure~\ref{fig:sim}), and our rank--based search algorithm is most
efficient when the dip in the lightcurve is isolated. Therefore, in
addition to searching for single--point events, we also bin our
lightcurves by 2, re-rank them and repeat the statistical tests
described above.  The probability of each data point is assessed for
both unbinned and binned lightcurves.  Each lightcurve is binned
twice, with two different starting points.  This increases the
detectability of occultations by large KBOs and by KBOs transiting
with low relative velocity.  For a detailed discussion of our
statistical analysis see \citet{TAOS_statpaper}.

For a set of lightcurves of a given star, the ranks in the three
telescopes should not be correlated for the statistical analysis
described above to be valid. We have developed a series of statistical
tests to identify data runs where significant correlations (typically
due to fast moving cirrus clouds) are found in the lightcurves. In
such data runs the ranks are not independently distributed and thus we
can not accurately determine the statistical significance of any
candidate events. Any data run where the independence of the
measurements after filtering cannot be rigorously established is
removed from our dataset.  A complete
description of these tests is beyond the scope of this paper, but they
are discussed in detail in \citet{TAOS_statpaper}
\begin{figure*}[ht!]
\centerline{\includegraphics[width=0.5\textwidth]{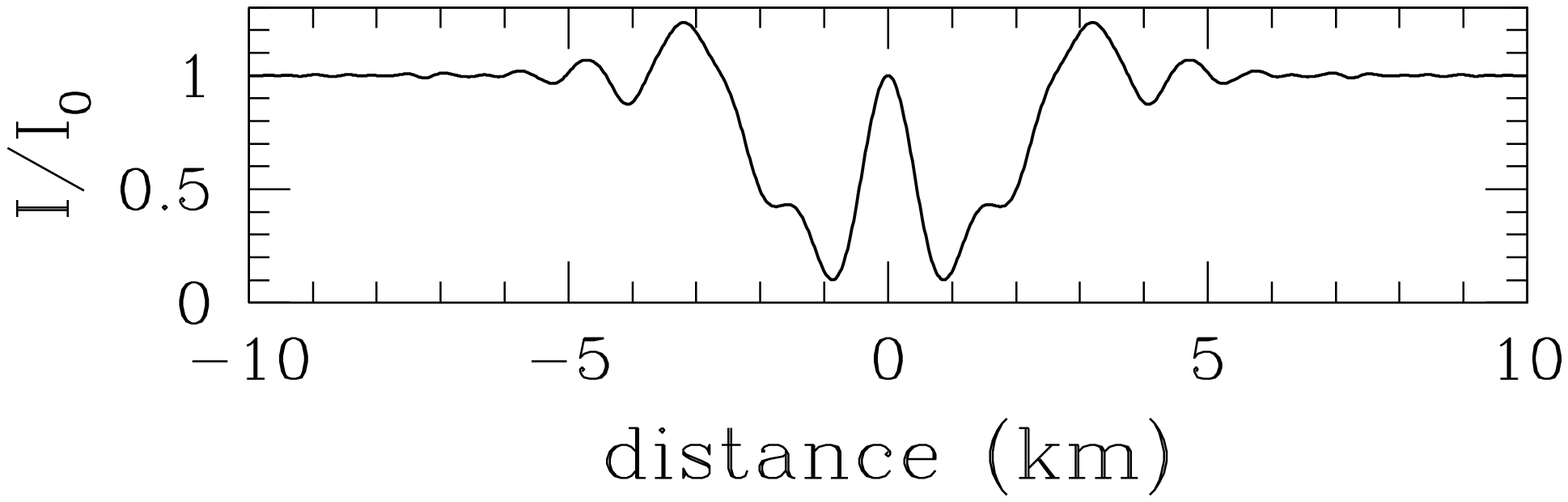}\includegraphics[width=0.5\textwidth]{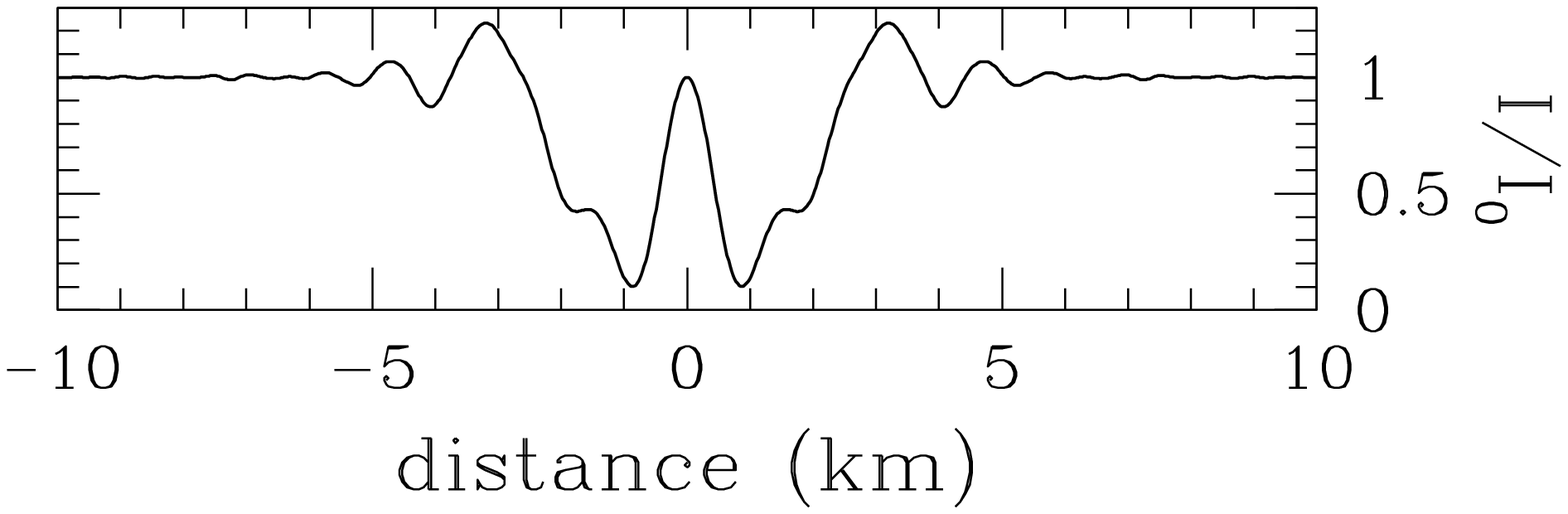}}
\centerline{\includegraphics[width=0.5\textwidth]{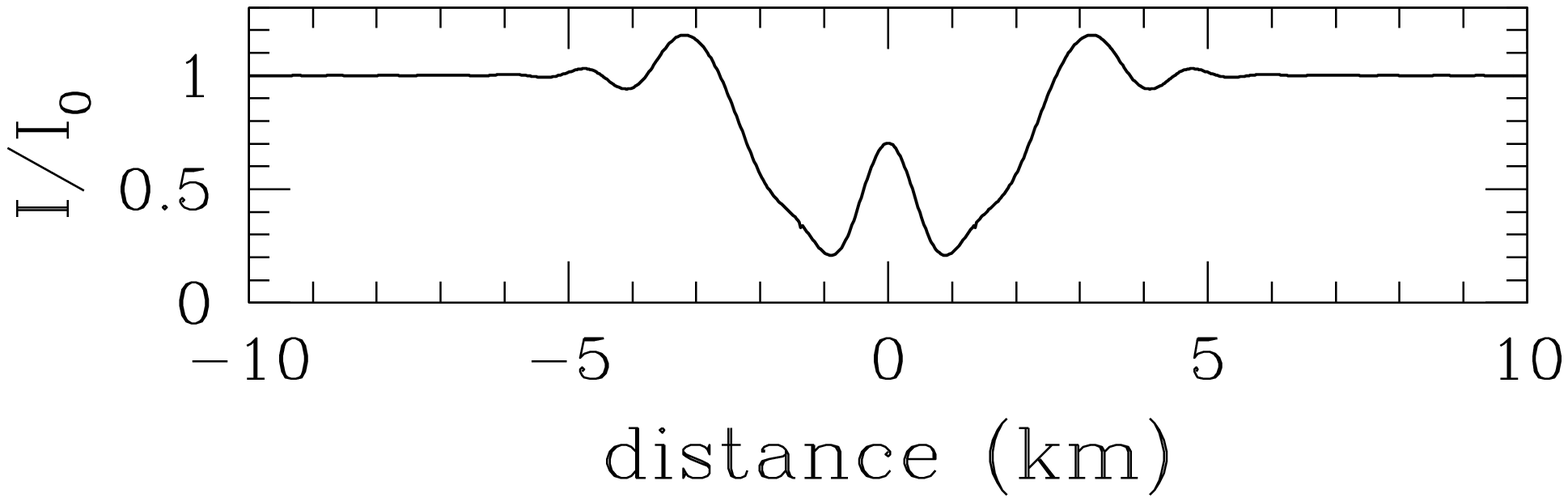}\includegraphics[width=0.5\textwidth]{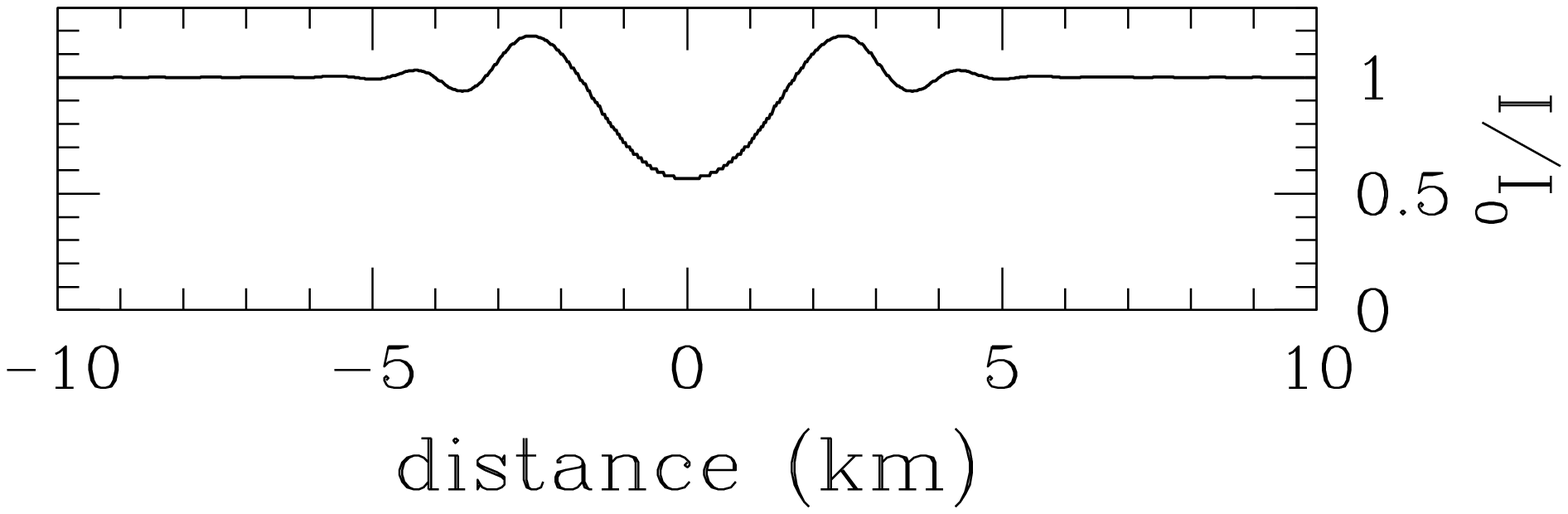}}
\centerline{\includegraphics[width=0.5\textwidth]{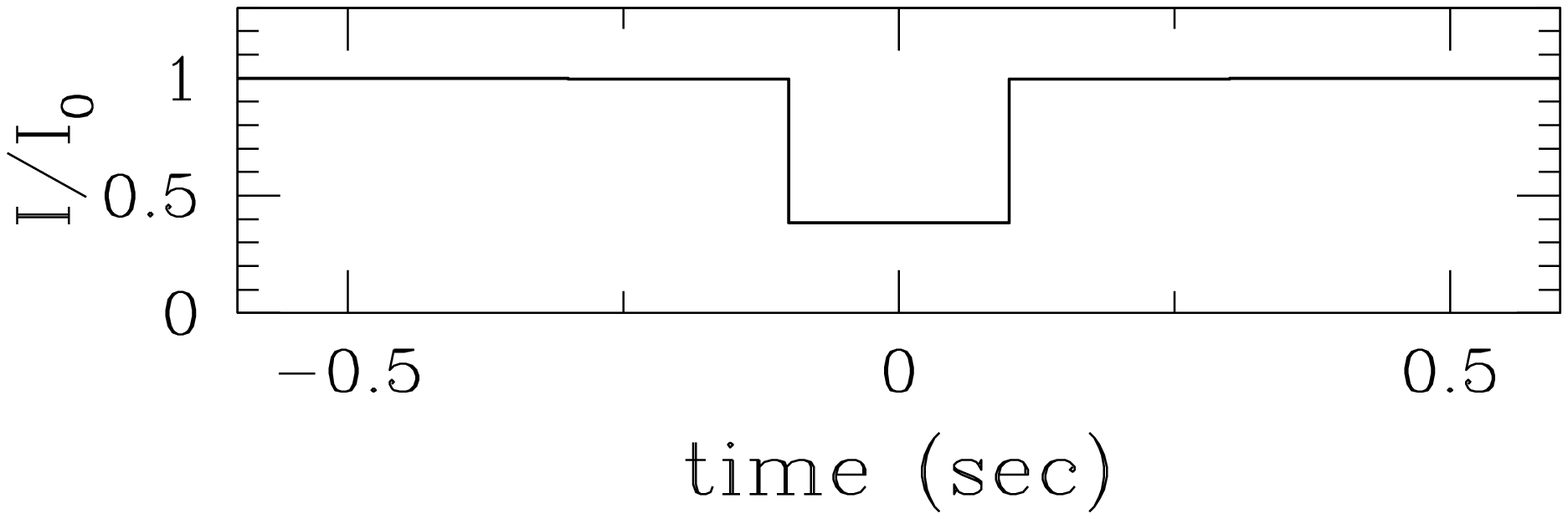}\includegraphics[width=0.5\textwidth]{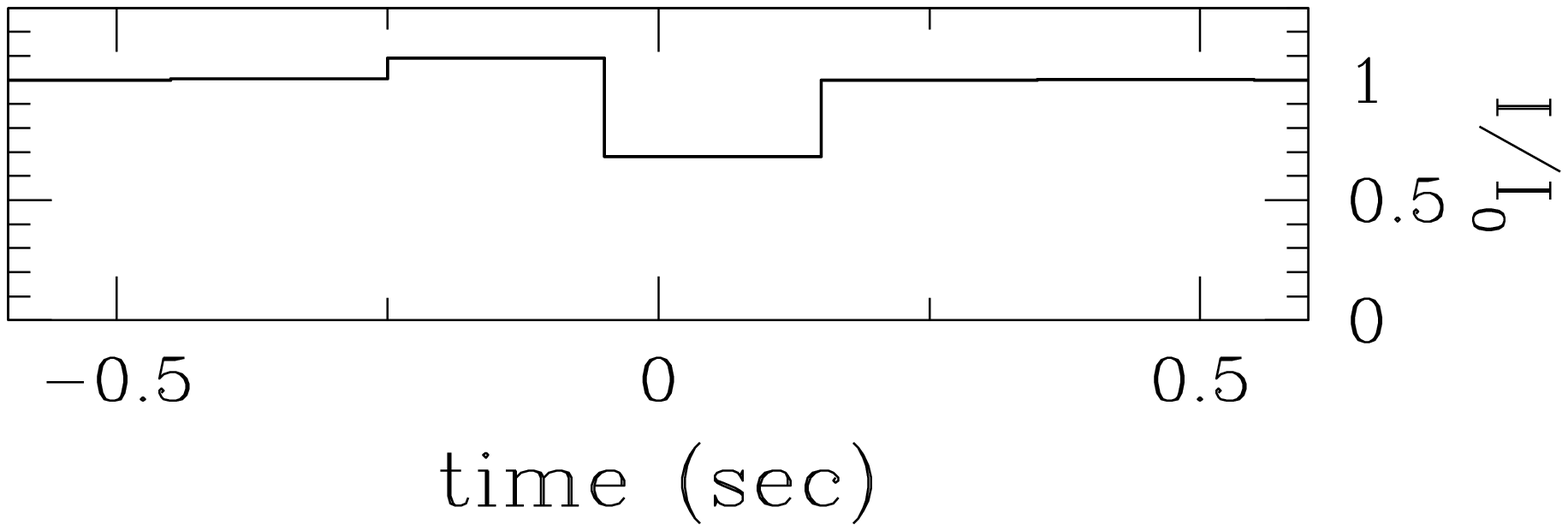}}
\caption{Steps of the generation of a simulated occultation
  event. Top, the left and right panels both show the same point
  source lightcurve for a 3~km KBO at 43~AU occulting an F0V
  star. Second row: finite source lightcurves for the same occultation
  parameters for a V~=~11 star, corresponding to an angular size of
  $0.015~\mathrm{mas}$, for a zero impact parameter (left), and at an
  impact parameter of 2~km (right).  Row three: the lightcurves in row
  two are integrated over intervals of 105~ms for the occultation
  above at opposition (left) and at $50\degree$ from opposition
  (right). The lightcurves are sampled at 5~Hz, with no time offset
  (left) and with a time offset of 50~ms (right).}\label{fig:sim}
\end{figure*}

For the next step we relax considerably the ultimate significance
requirement described above, and select as \emph{provisional
  candidates} those triplets that have $\xi \leq
1.0\times10^{-6}$. Note that the significance $\xi$ refers to the
probability that the point would be drawn from a random distribution,
therefore the lower the value of $\xi$ the higher the statistical
significance of the event.  Nearly 150,000 provisional candidates are
found.  We use all these measurements to identify and remove spurious
regions of the lightcurves, and hence identify and reject false
positive events which arise from sources other than random chance.
The constraints described below allows us to recognize regions of
lightcurves with atypical noise and contamination by transiting
objects (satellites or meteors, which turn out to be the major source
of false alarms), and to identify high-frequency fluctuations in the
raw data that are not removed by the high-pass filter. These are the
steps of our false positive rejection process:
\begin{itemize}

\item \emph{Contiguity:} Contiguous candidates within a lightcurve and
  candidates that are within three time-stamps of each other are
  removed. Only the one rank triplet that has the highest significance
  in a series of contiguous or proximate points is considered as a
  candidate. This removes double--counted events: events caused by
  large KBOs or KBOs moving at low relative velocity would affect more
  than one contiguous point. Furthermore this removes events that are
  double--counted because they appear significant in both the binned
  and unbinned lightcurves. This eliminates about 40\% of the
  candidates.

\item \emph{Simultaneity:} Candidates that appear in the lightcurves
  of more than one star simultaneously at the same time-stamp
  \emph{or} within three rowblocks are considered to be false
  positives. We expect simultaneous count drops in time-domain to be
  primarily due to inaccurate aperture positioning in the
  photometry. In the rowblock domain simultaneous count drops might be
  due to inaccurate background determination or non-occultation events
  altering the baseline of the lightcurve at, or around, the candidate
  event, or by fast moving cirrus clouds or other phenomena which
  induce high frequency fluctuations in multiple lightcurves which are
  not removed by the high-pass filter.  This cut removes about 60\% of
  the remaining candidates.

\item \emph{Number of telescopes:} At this point we require all of our
  remaining candidates to have been observed by all three telescopes.
  Although we are only considering three-telescope runs in our
  analysis, for some targets the lightcurve might not be extracted in
  the photometry phase for all telescopes. Small differences in the
  field of view and in the field distortion might make one star target
  not visible to all telescopes if it is at the edge of the field or
  if the crowding induced by the zipper--mode readout caused overlap
  of the target with other stars~\citep{2008arXiv0802.0303L}.  About
  35\% of the remaining candidates are thus removed. Note that this
  cut cannot be applied earlier as simultaneous events might appear in
  only one 3--telescope lightcurve, but also in 2--telescope
  lightcurves, and we want to be able to recognize and remove these
  events.  At this point there are still over 20,000 candidates
  left. Note that we do not count the star hours discarded by this cut
  in our total exposure of $5 \times 10^5$ star--hours.

\item \emph{Significance threshold:} We finally constrain $\xi$ such
  as to expect fewer than 0.27 false positives in our dataset.  This
  constraint depends on the size of the dataset: for the $9\times10^9$
  triplets remaining $\xi < 3.0\times10^{-11}$ allows $<0.27$ false
  positives due to random noise. Only 228 candidates remain.
\end{itemize}

The remaining candidates require visual inspection: first of the
lightcurves, and for any remaining candidates, of the images. Most of
the events are caused by the passing of bright objects, such as
artificial satellites, meteorites or asteroids, that generate a
variation in the background or baseline of the lightcurve responsible
for causing artificially low counts in the neighborhood of the
object. Many, but not all, of these false positives are removed by the
simultaneity cut described above. Note that in our observing mode
bright stars generate a bright streak across the length of our images
(see~\citealt{TAOS_photpaper}), as flux is collected during the
shutterless row shifting. In the presence of a bright object
overlapping with a star--streak generated by the zipper--mode readout,
the brightness of the streak is overestimated, thus too much flux is
subtracted from the rowblock column causing an artificial flux drop in
the star time series. In many instances the foreground object will
also appear inside the star aperture artificially boosting its
brightness. This flux drop will then be associated with a very high
flux measurement following or preceding the event epoch, a signature
that allows us to remove these false positives by inspecting the
lightcurve.  We also inspect the centroid position of the aperture. If
the aperture position has moved significantly at the time-stamp of the
candidate the candidate is rejected. Of the remaining candidates, 90\%
are rejected by visual inspection of the time series.

Finally we inspect the \emph{images} of the remaining 23 candidates:
they also were all associated with bright moving objects overlapping
star streaks. No candidate events were left in our dataset at the
conclusion of this process.

\subsection{Determination of the Stellar Angular Size}\label{sec:angs}

The shape of a lightcurve during an occultation event, and hence the
detection efficiency, is strongly dependent on the angular size of the
target star (\citealt{2000Icar..147..530R},
\citealt{2007AJ....134.1596N}, \citealt{2009arXiv0902.3457B}).  Our
fields contain a variety of stellar types and a large range of angular
sizes (Figure~\ref{fig:angsz}a). Here we describe the method we use to
estimate the angular sizes of our target stars in order to account for
this effect when estimating our detection efficiencies (see Section
\ref{sec:analysis}).

Angular sizes have been related to the position of a star in the
color-color or color-magnitude diagrams (e.g.,
\citealt{1999PASP..111.1515V}, \citealt{2002AJ....123.3380N}).  We
follow the work of \citet{2002AJ....123.3380N} and calculate the
angular size of our star targets using the 2MASS $J$ and $K$
color~\citep{2003tmc..book.....C} to invert the set of equations:
\begin{eqnarray}
F_K &=& (3.942 \pm 0.006) - (0.095 \pm 0.007) (J-K)\label{eq:sb}\\
F_K &=& 4.2207 - 0.1K - 0.5 \log\theta \label{eq:angsoumen}
\end{eqnarray}
where $F_K$ is the surface brightness of a star in K-band, which is
related to its $J-K$ color, as well as to its unreddened
apparent K magnitude and angular size $\theta$.  The
relationship between the surface brightness and the color of a star
(Equation~\ref{eq:sb}) is calibrated using angular sizes measured
directly by long baseline interferometry~\citep{2002AJ....123.3380N}.

Not all of our target stars, however, are identified 2MASS objects,
while in the photometry phase we have identified all of our targets
with USNO-B objects.  We therefore devised a method that relies on
USNO-B $R$ and $B$ magnitude to calculate the angular sizes of our
targets.

We first derive the angular size of a subset of targets identified
with 2MASS objects using the above equations, and scale it to obtain
the angular size the targets would have if their apparent magnitude
were $R~=~12$. We then considered the USNO-B $B-R$ color for all of
these targets and calculated a regression on these points.  This
generates a formula that allows us to go from the USNO-B color of any
of our targets to 2MASS colors and thus predict angular sizes
according to Equations~\ref{eq:sb}~\&~\ref{eq:angsoumen}, for an
apparent magnitude $R=12$. To calculate the true angular size we
rescale from $R=12$ to $R_\mathrm{USNO}$\footnote{We do not use our
  instrumental magnitude for rescaling for consistency with what is
  used in the color determination.}.  The angular sizes of a subset of
TAOS targets, rescaled to $R=12$, is plotted as derived from
Equation~\ref{eq:sb}~\&~\ref{eq:angsoumen} versus the USNO-B $B-R$
color (Figure~\ref{fig:angsz}b). Our regression on the data is plotted
as well (solid line).

The scatter in the determination of the angular size via the method
described above is large, as can be seen in
Figure~\ref{fig:angsz}b. This is due to scatter in the USNO-B color
($\approx 0.3~\m{mag}$,~\citealt{2003AJ....125..984M}), to the (much
smaller) scatter in the $J$ and $K$ magnitudes, and to the scatter in
the empirical determination of the relationship between $\theta$ and
$J-K$ in Equation~\ref{eq:sb}~\&~\ref{eq:angsoumen}.  We have not used
any interstellar reddening corrections, and the angular size
estimation of an unknown reddened star from the near-IR relationship
would be relatively less affected compared to that in visual bands.
Reddening is typically small for our targets though, since we are only
considering objects brighter than $R\sim13.5$.

The distribution of angular sizes is well
reproduced. Figure~\ref{fig:angsz}a shows the distribution of angular
sizes for a typical TAOS field, calculated via
Equations~\ref{eq:sb}~\&~\ref{eq:angsoumen}, and
Figure~\ref{fig:angsz}c shows the distribution of angular sizes in our
efficiency simulation obtained via the USNO-B color. The distributions
do overlap. About 2\% of our simulated angular sizes fall in the
region $\theta>0.15~\mathrm{mas}$ and $R_\mathrm{TAOS} > 11$, where
there are no observed objects. These objects have poor USNO-B color
determination. Figure~\ref{fig:angsz}d  shows the region of the
$\theta-R_\mathrm{TAOS}$ space where simulated events \emph{are
  recovered}. There are few recoveries in the region
$\theta>0.15~\mathrm{mas}$ and $R_\mathrm{TAOS} > 11$, so these stars
do not contribute the the expected event rate.

\subsection{Detection Efficiency}\label{sec:analysis}

It is necessary to assess the efficiency of our recovery algorithm in
order to derive the number density of KBOs from the number of events
in our survey.  In order to measure our recovery efficiency we implant
our data with synthetic occultations.  The data are then reprocessed
in the same way we did to search for true events.  By implanting into
the actual lightcurves we do not make any assumption regarding the
nature of the noise in our data.  Note that our detection algorithm,
described in Section~\ref{sec:fp}, is not affected by the spectral
characteristics of the noise, as long as the distribution of flux
measurements in a lightcurve is stationary~\citep{TAOS_statpaper}. Our
occultation simulator is based on the work described in
\cite{2007AJ....134.1596N}.
\begin{figure*}[t!]
\centerline{\includegraphics[width=0.4\textwidth]{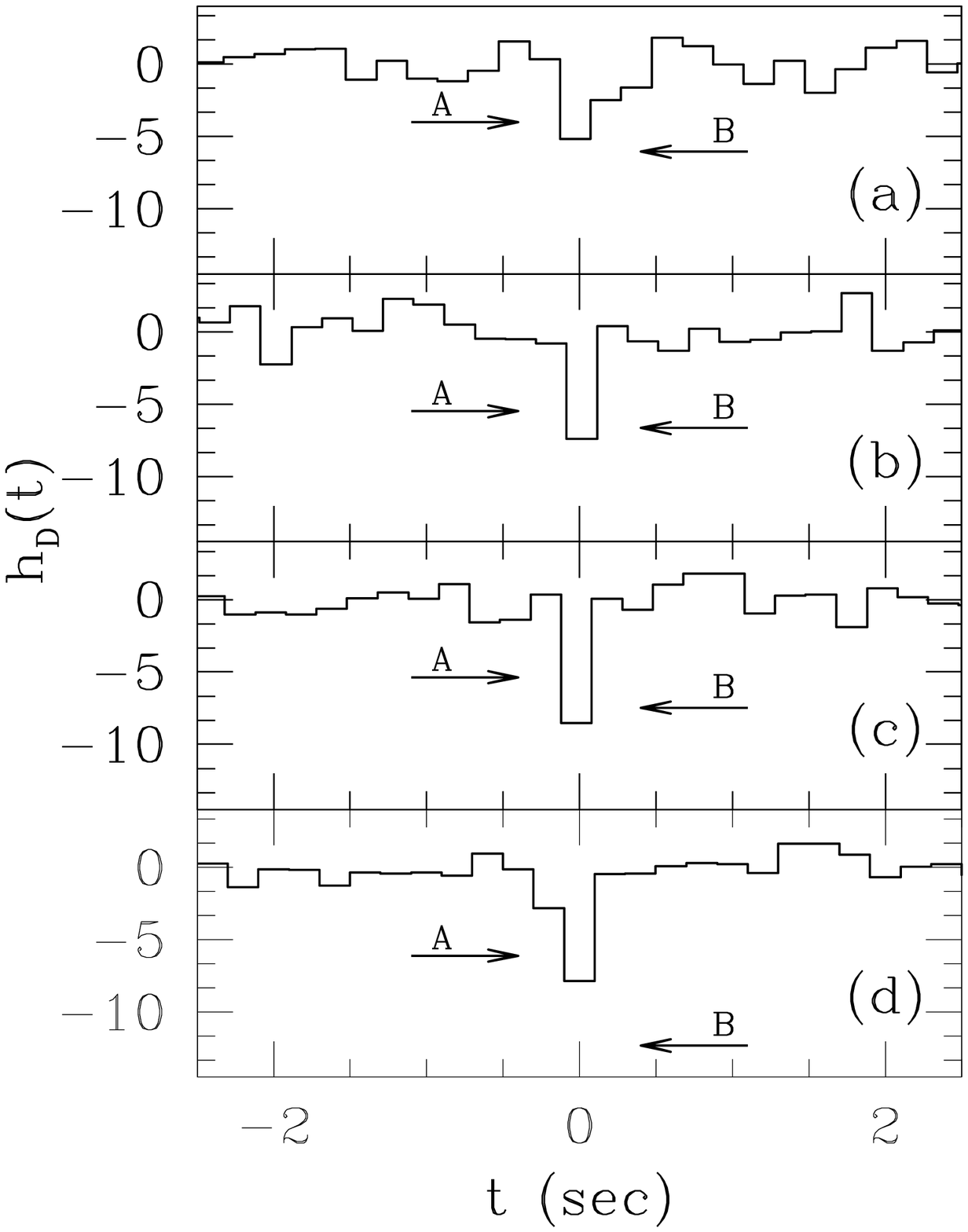}\includegraphics[width=0.4\textwidth]{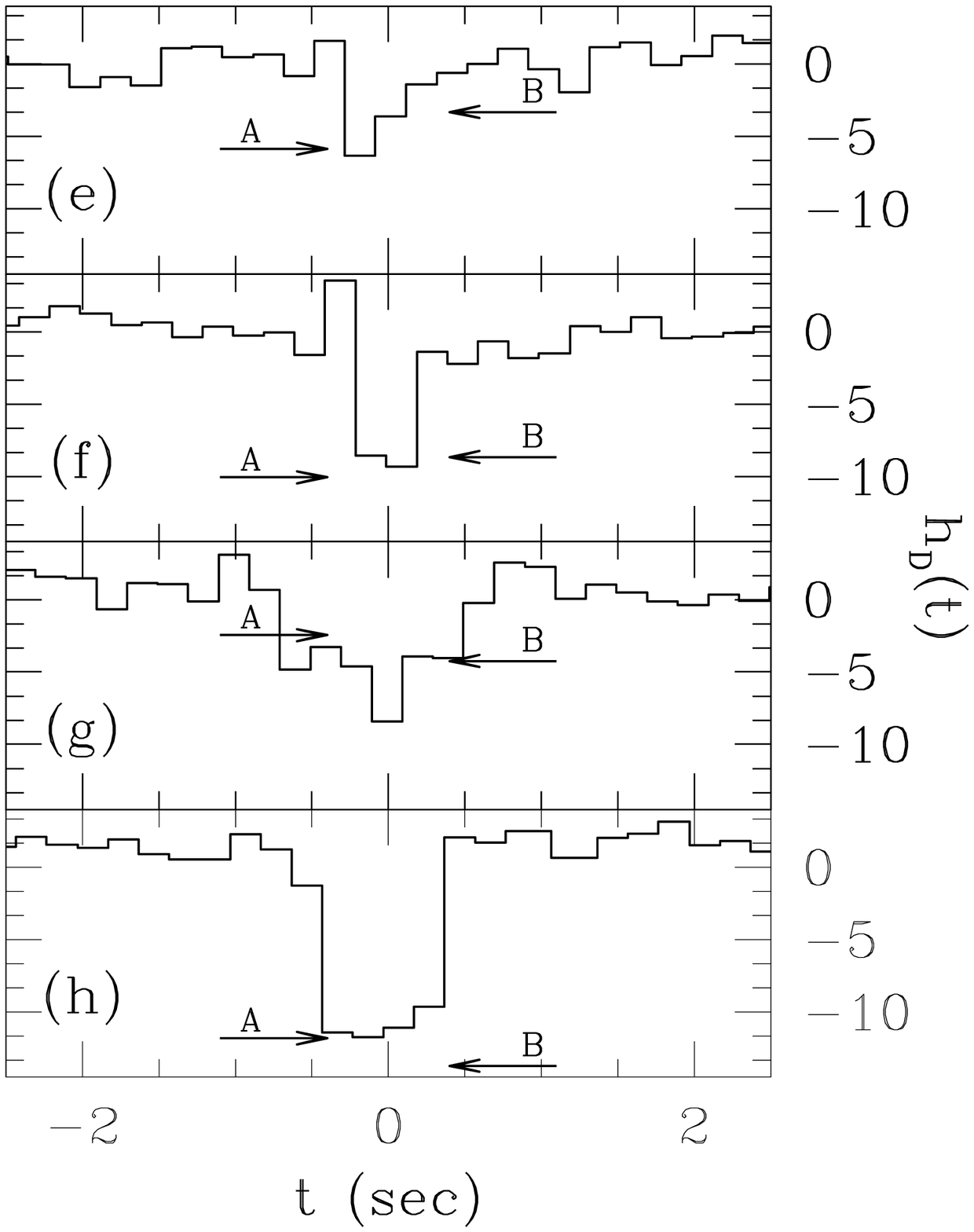}}
\caption{Implanted occultations recovered by our pipeline. On the left
  the lightcurves refer to star targets in a field observed near opposition
and on the right in a field observed at large angle from opposition.
deviation at that epoch from the local mean of the lightcurve in units
of standard deviation as measured by telescope TAOS D. The value of
$h_\m{A}(0)$ and $h_\m{B}(0)$, for TAOS A and B, are indicated by
horizontal arrows labeled A and B, respectively.  Each event is
described in Table ~\ref{tab:fakeocc}}\label{fig:fakeevt}
\end{figure*}

\begin{deluxetable}{lcc}[b!]
\tablecolumns{3}
\tablewidth{0pc}
\tablecaption{Distribution of synthetic events}
\tablehead{diameter (km)   & implantations  & recoveries }
\startdata
30.0    & 231 &    75\\
8.0    & 385 &    84\\
3.0    & 1078 &    73\\
2.0    & 2003 &    89\\
1.3    & 4393 &    73\\
1.0    & 13255 &    66\\
0.7    & 36222 &    40\\
0.5    & 447764 &    9
\enddata
\label{tab:implant}
\end{deluxetable}
\begin{deluxetable*}{lcccccccc}[t!]
\tablecolumns{11}
\tablewidth{0pc}
\tablecaption{Parameter of implanted events in Figure~\ref{fig:fakeevt}}
\tablehead{ &$D$~(km)&$b$~(km)&$v_\m{rel}$~(km/s) & SNR &$\theta_\star$~(mas) &$h_\m{A}(0)$ &$h_\m{B}(0)$ &$h_\m{D}(0)$}

\startdata
(a)&0.7 & 0.50 & 25.4 & 39.8 & 0.03 & -4.0 & -6.1 & -5.2\\
(b)&1.0 & 0.50 & 14.9 & 29.4 & 0.03 &-5.5 & -6.6 & -7.4\\
(c)&3.0 & 1.19 & 25.4 & 12.8 & 0.03 & -5.4 & -7.5 & -8.5\\
(d)&8.0 & 2.7 & 25.4 & 9.4 & 0.006 & -6.1 & -12.3 & -7.9\\
(e)&0.7 & 0.22 & 8.2 & 10.2 & 0.03 & -5.8 & -3.3 & -6.3\\
(f)&1.0 & 0.94 & 7.7 & 10.8 & 0.03 & -10.0 & -8.7 &-8.6\\
(g)&3.0 & 1.18 & 3.1 & 12.9 & 0.006 & -2.5 & -4.2 & -8.4\\
(h)&8.0 &2.93 & 8.1 & 15.7 & 0.04& -11.8 & -13.7 & -11.7
\enddata
\label{tab:fakeocc}
\end{deluxetable*}

We first generate diffraction lightcurves for KBOs occulting point
sources. We integrate the diffraction pattern over the disk of our
target star. Keeping the stellar type fixed, the angular size is
modulated by changing the apparent magnitude of the star and we can
use the point source lightcurve to integrate the occultation signature
over the star disk.  A point source lightcurve for a $D=3~\mkm$ KBO at
$\Delta=43$~AU is shown in Figure~\ref{fig:sim}, top, and the finite
source lightcurve for a magnitude $V=11$ star is shown in the second
panel, left. Note the smoothing of the diffraction features.  We
modify the lightcurve to account for a finite impact parameter $b$ by
using the finite source lightcurve at impact parameter $b=0$ as an
input and calculating the intensity of the occultation signal at the
new distance of each point form the center of the diffraction pattern
by interpolating points of the finite source lightcurve.  A lightcurve
for an F0V, $V$=11 star and a 3~km KBO occulting at an impact
parameter $b~=~2~\mkm$ is plotted on the right hand side of the second
panel of Figure~\ref{fig:sim}.  Finally, after calculating the
relative velocity of the KBO as a function of distance and angle from
opposition as per~\cite{2004Liang}, and~\cite{2007AJ....134.1596N}, we
smooth the lightcurve to account for finite exposure intervals, and we
sample the \emph{finite exposure} lightcurve at the appropriate
sampling rate. In this step we can account for dead-time in the
sampling interval, which for TAOS is $47.5\%$ at 5~Hz. We also allow
an offset in time between the center of the finite sampled lightcurve
and the integration bin. In the bottom row of Figure~\ref{fig:sim} the
lightcurves for the event in row two are integrated over 105~ms second
intervals and sampled at 5~Hz, the typical sampling rate of TAOS, for
an event at opposition and with no time offset (left), and for an event at
$50\degree$ from opposition with an offset of 50~ms between the center
of the sampling interval and the center of the occultation (right).

\begin{figure*}[ht!]
\centerline{\includegraphics[width=0.35\textwidth]{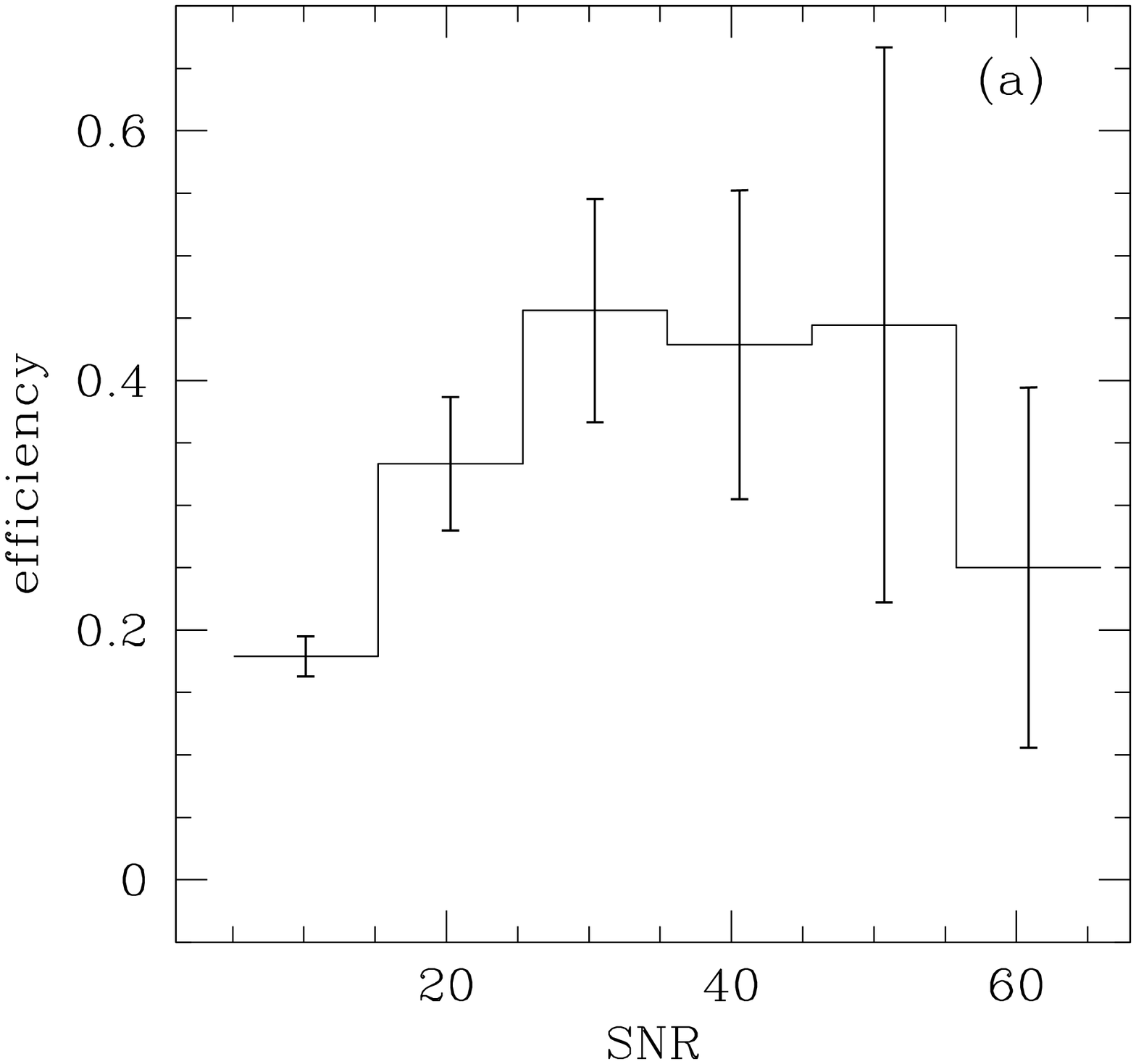}\includegraphics[width=0.35\textwidth]{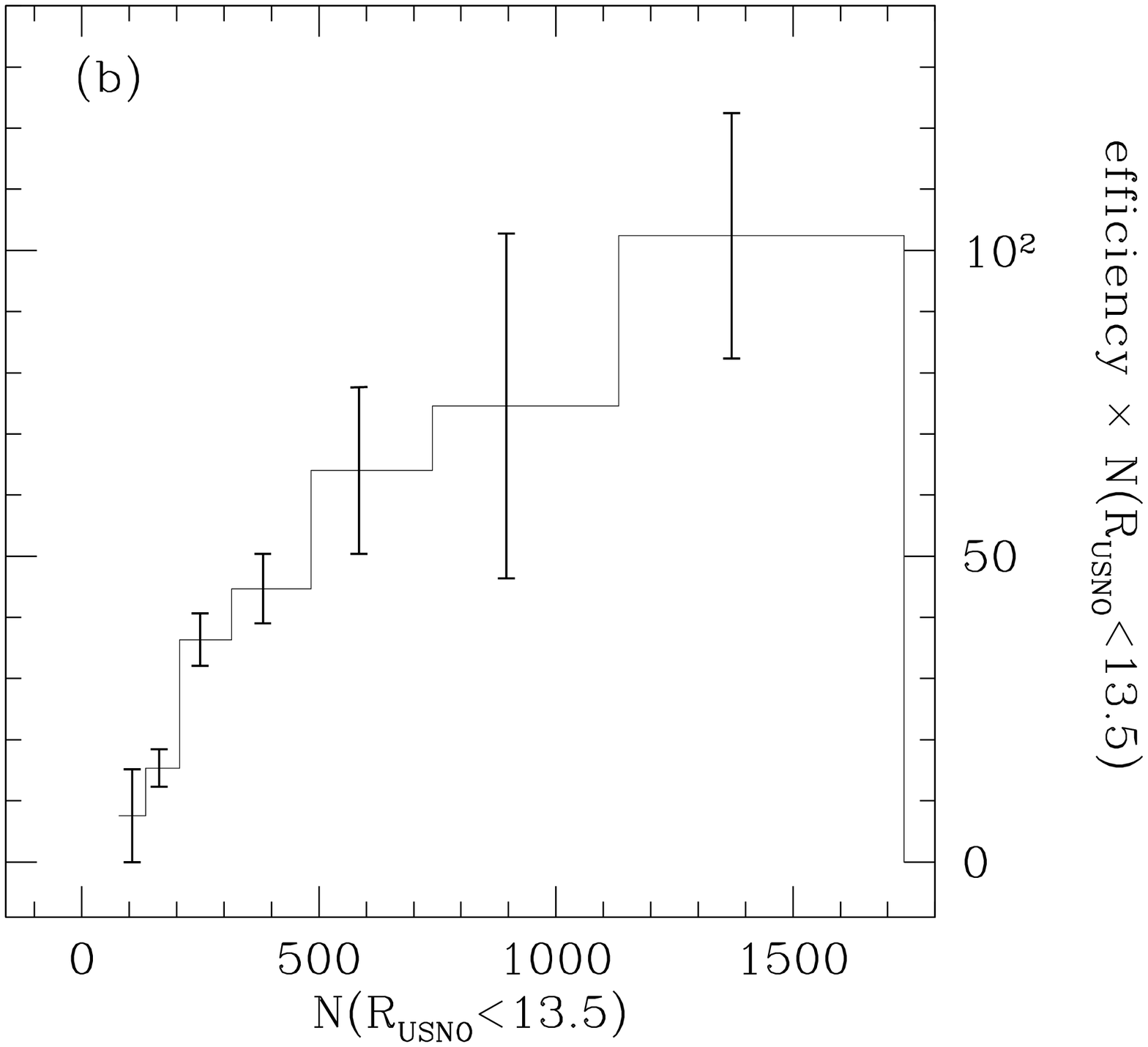}}
\centerline{\includegraphics[width=0.35\textwidth]{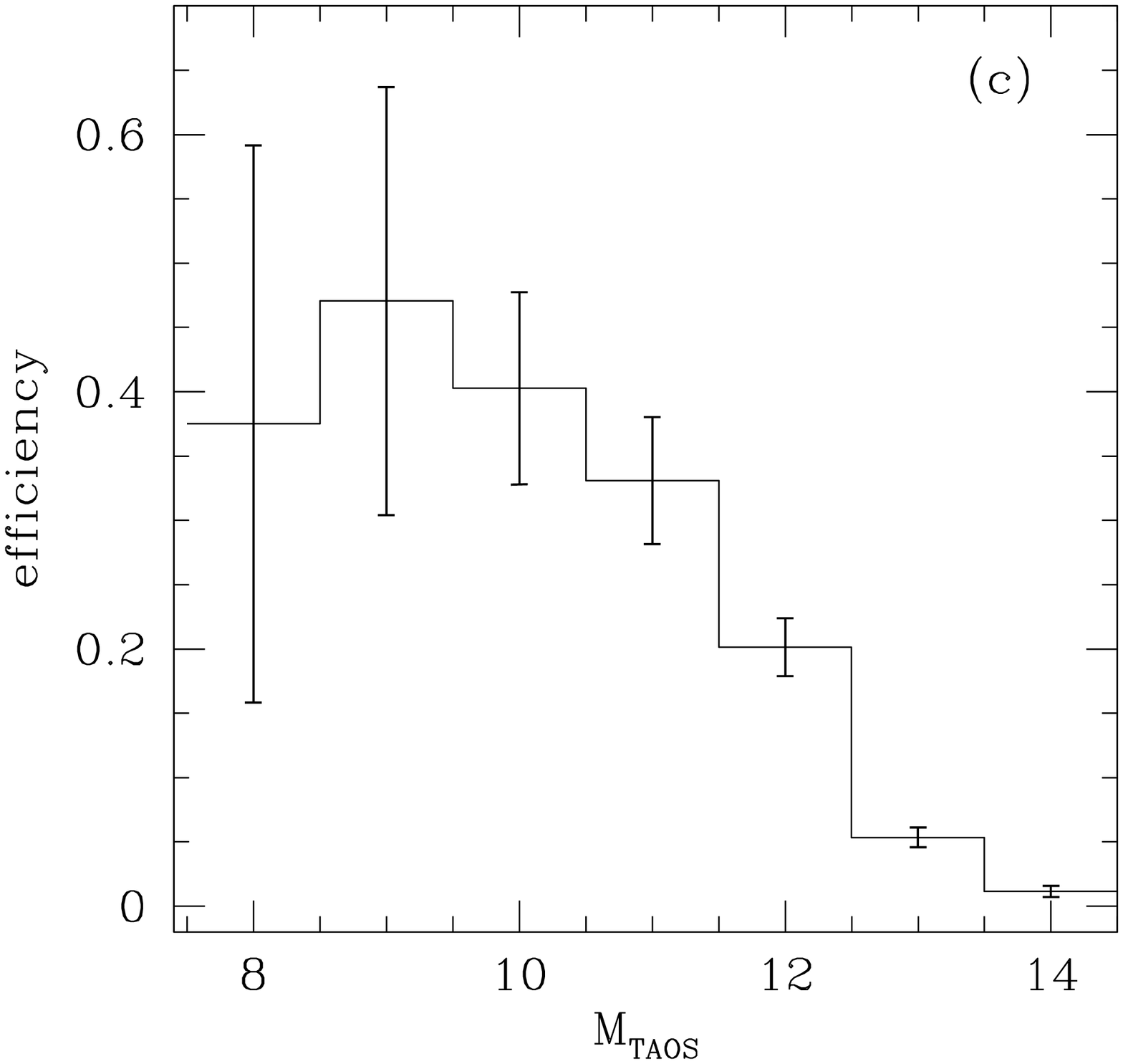}\includegraphics[width=0.35\textwidth]{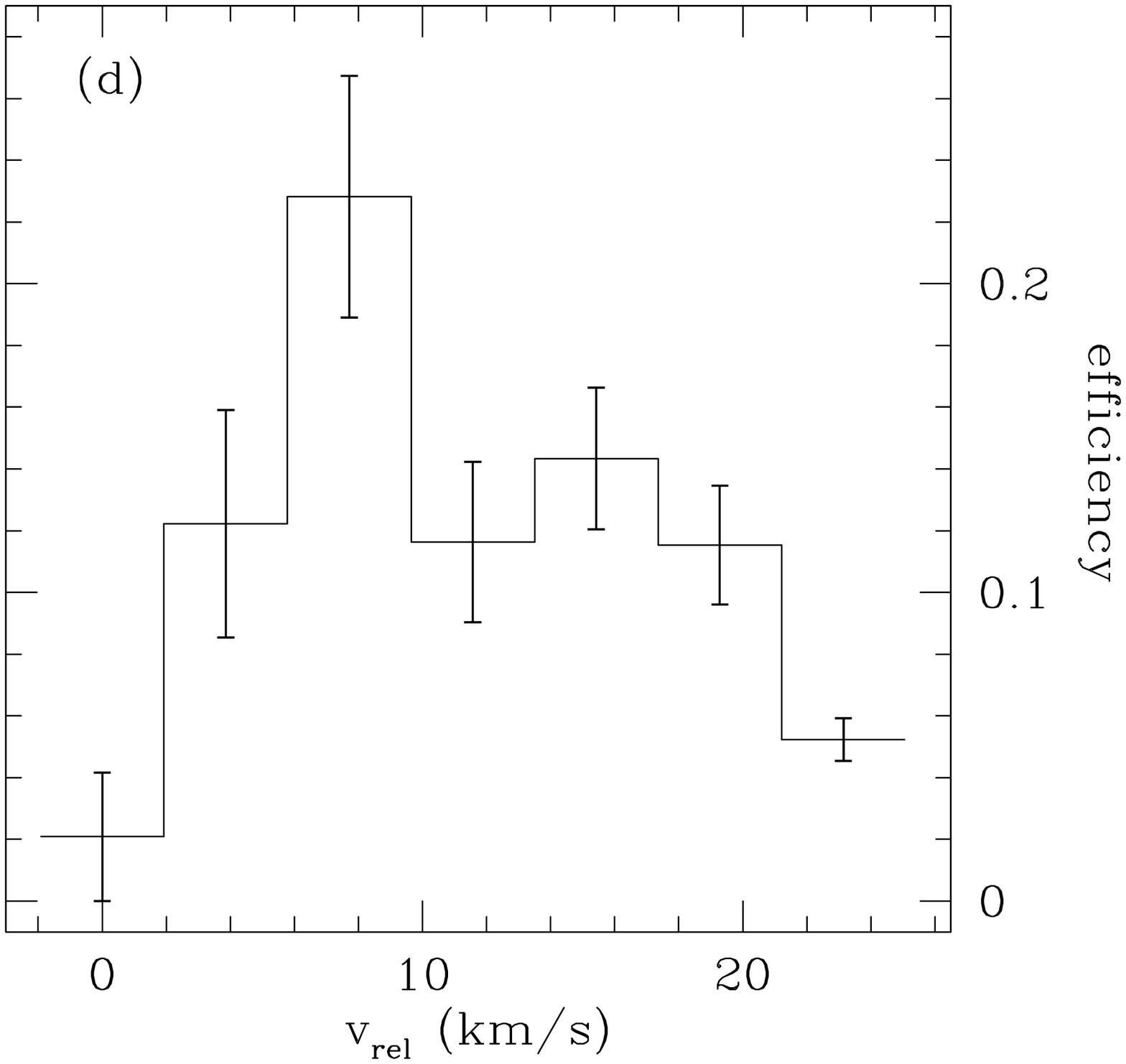}}
\centerline{\includegraphics[width=0.35\textwidth]{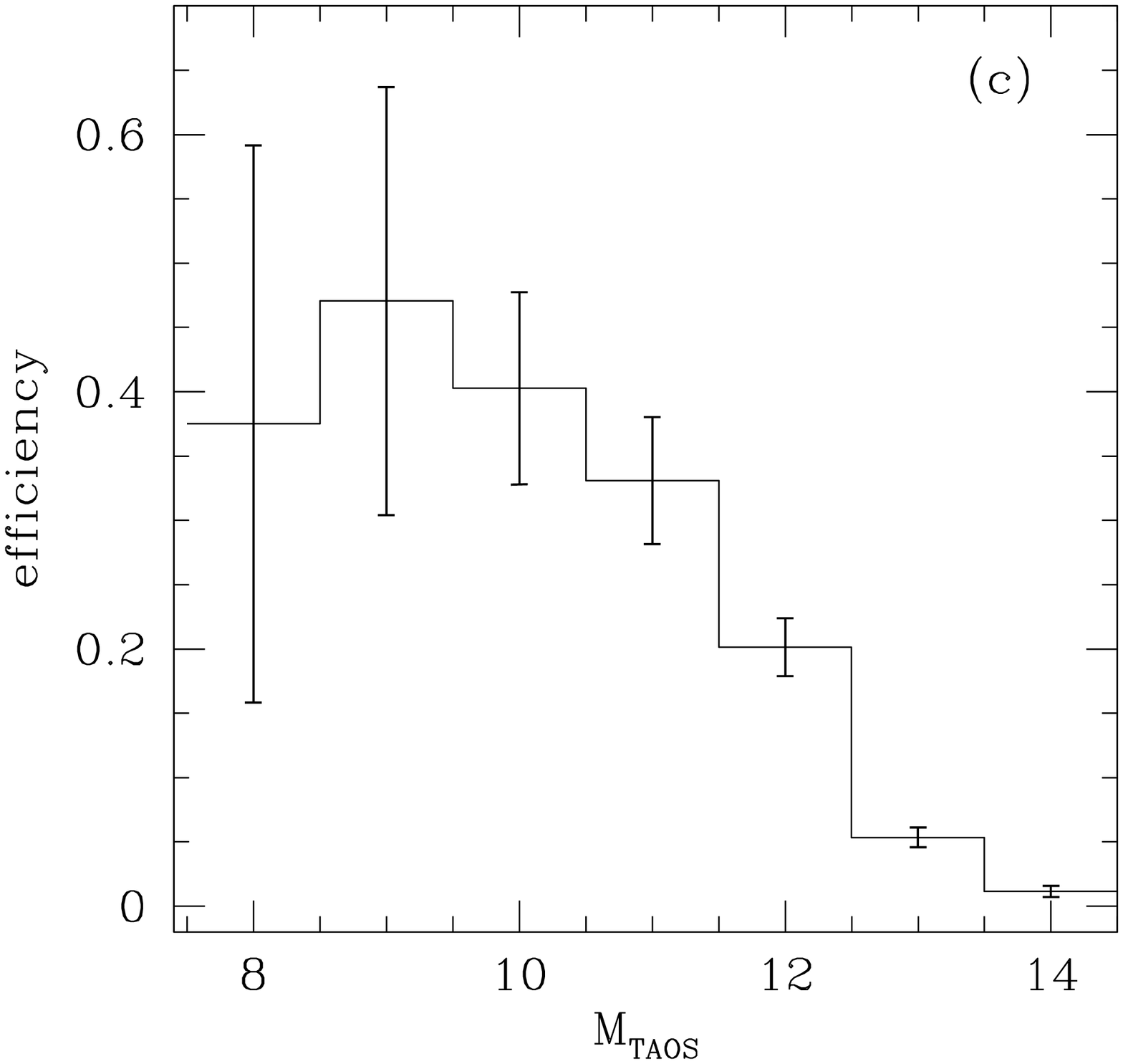}\includegraphics[width=0.35\textwidth]{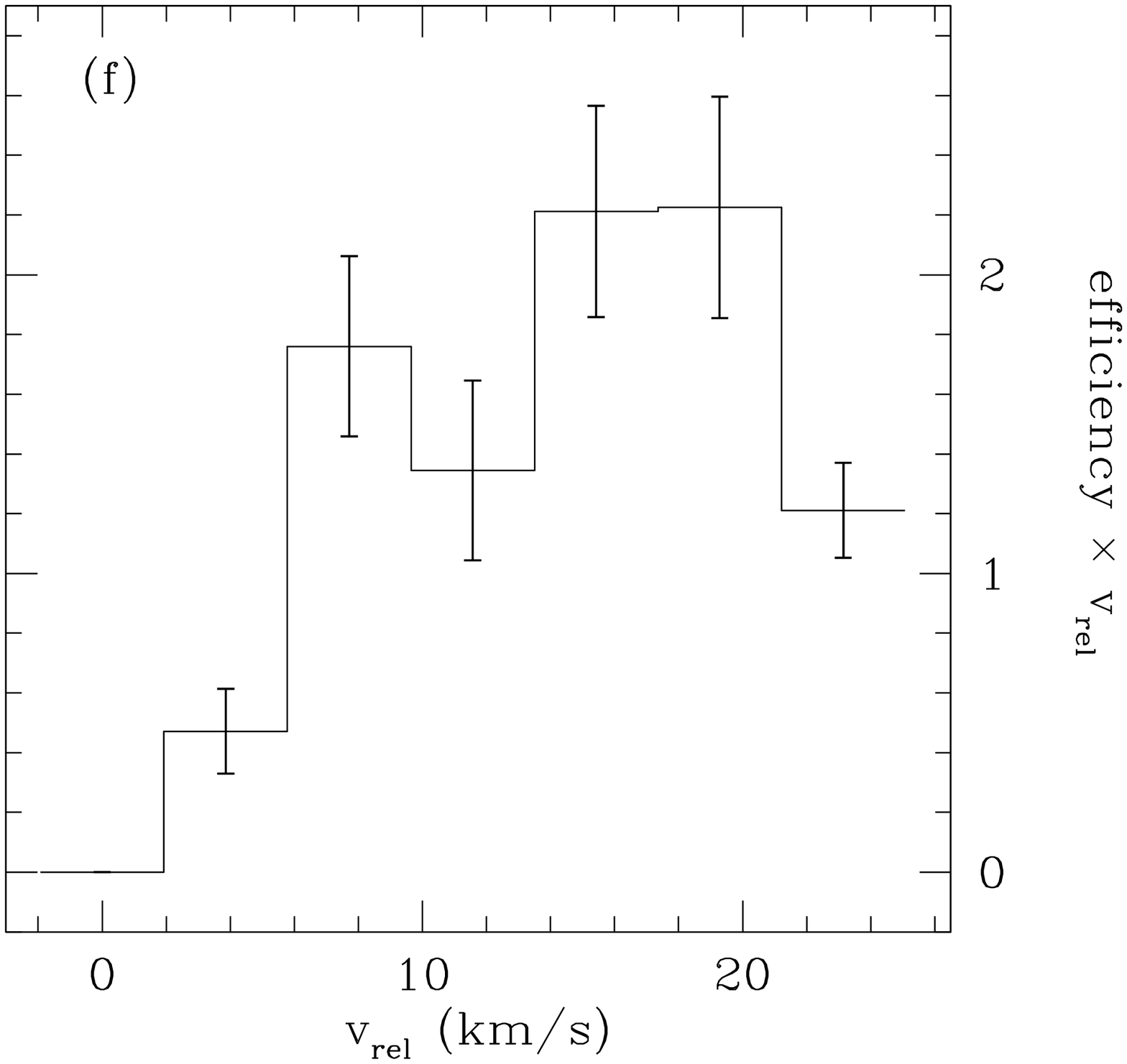}}
\caption{Recovery efficiency for 3~km KBOs. Panel (a): efficiency as
  function of SNR. Efficiency versus crowdedness of the field in panel
  (b), defined as the number of targets brighter than $R_\m{USNO} =
  13.5$. The efficiency is weighted by the crowdedness.  Panel (c):
  efficiency versus magnitude $M_\m{TAOS}$ . In panel (e) the
  efficiency as a function of magnitude is weighted by the number of
  targets at that magnitude.  Panel (d): efficiency versus relative
  velocity of the KBO targets. In panel (f) the efficiency versus
  relative velocity is weighted by the relative velocity. All error
  bars are calculated in a Poissonian fashion from the square root of
  the number of recoveries.}\label{fig:eff_vcrowd}
\end{figure*}


In order to sample properly the space of diameters to which the survey
is sensitive we implant synthetic occultations by objects of diameter
$D=0.5$, 0.7, 1.0, 1.3, 2.0, 3.0, 8.0, and 30.0~km. For a 30~km
diameter KBO the event falls in the geometric regime, diffraction
effects are therefore no longer significant and our efficiency
stabilizes. Because our sensitivity decreases with decreasing diameter
we implant progressively more objects at smaller diameters. The number
of implantations at each size is designed to allow us to obtain a good
sampling at all sizes. In Table~\ref{tab:implant} we report the number
of objects implanted for each size in one of our efficiency runs, and
the number of recoveries\footnote{Four runs are conducted to improve
  statistical accuracy in the determination of our efficiency.}. For
objects within the Kuiper Belt (about 30 to 60~AU), the differences
induced by different distances are negligible in the occultation
features as observed by TAOS. We therefore set the distance to
$\Delta~=~43~\mathrm{AU}$.
Every occultation event is implanted at a random epoch in the
lightcurve set and at a random impact parameter between 0 and $H/2$,
where we set $H$, a measure of the cross section of the event, to the
size of the Airy ring and the projected size of the star in accordance
to \citet{2007AJ....134.1596N}.

In order to implant the synthetic occultations into our data we
modulate the lightcurve by subtracting (adding) the amount of flux
suppressed (augmented) by the occultation at each data-point as done
in Z08.  This approach slightly overestimates the noise due to Poisson
statistics where the flux is suppressed, giving us a conservative
estimate of our efficiency. We implant exactly one occultation in each
lightcurve in our dataset.

\begin{figure}[t!]
\centerline{\includegraphics[width=0.5\textwidth]{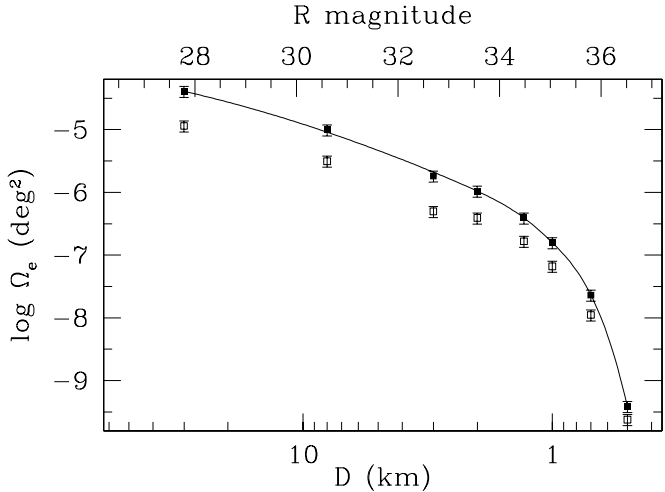}}
\caption{Effective solid angle for the first 2 years of TAOS data (Z08, empty squares) and for the current 3.75 year dataset (solid line and filled squares).}\label{fig:Omegas}
\end{figure}

\begin{figure}[b!]
\centerline{\includegraphics[width=0.5\textwidth]{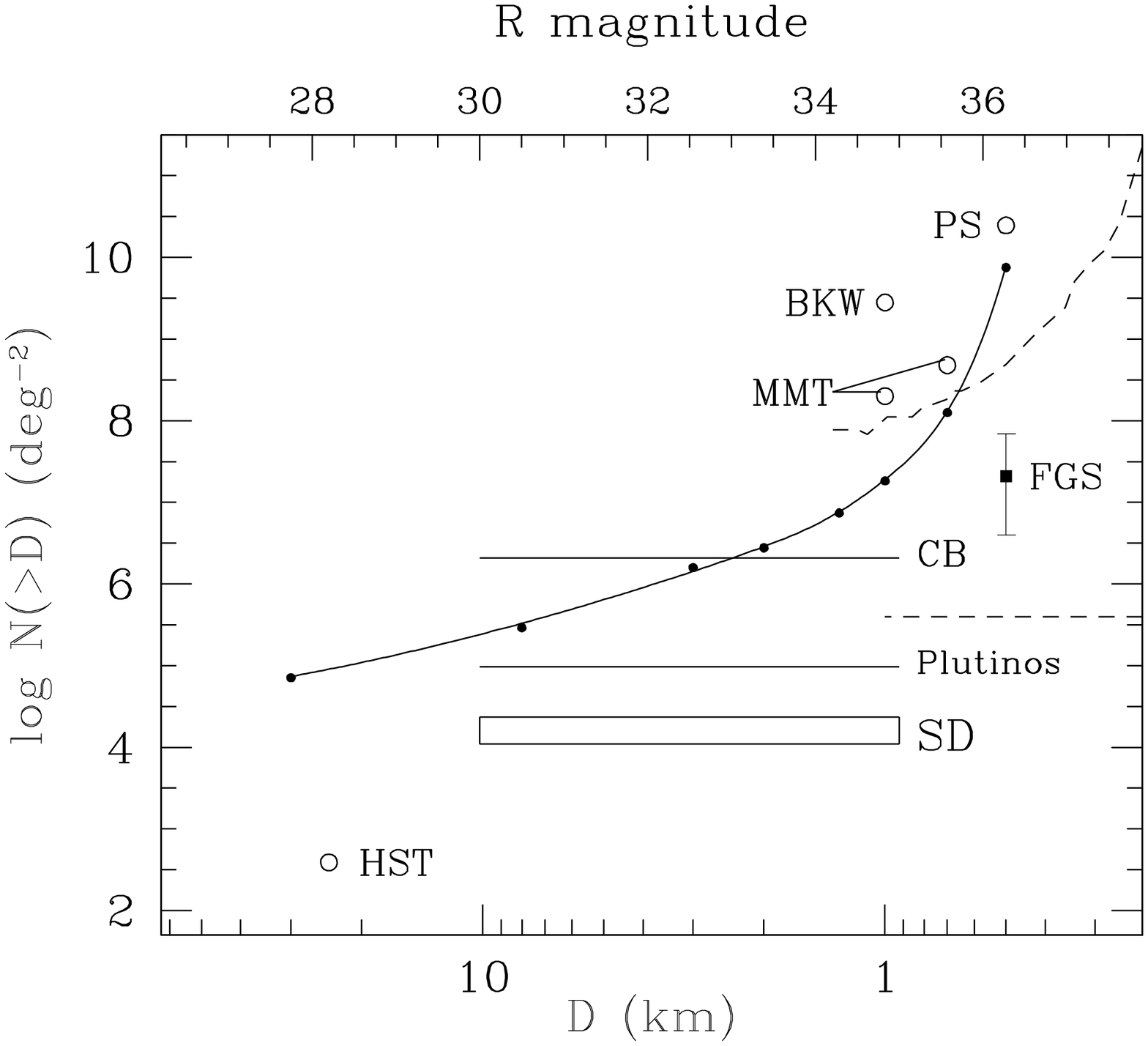}}
\caption{Model-independent upper limits from the TAOS survey (solid
  dots and solid line). Each point represents the upper limit to the
  number of KBO of that size or larger given the TAOS effective
  coverage.  Similarly derived upper limits from
  \citet{2008AJ....135.1039B} at 1~km (BKW), \citet{ps} at 0.5~km (PS)
  and \citet{mmt} at 1~km and 0.7~km (MMT) are plotted as empty
  circles.  The upper limit from~\citet[direct
    survey]{2004AJ....128.1364B} is also plotted (HST). The dashed
  lines represent the 95\% upper and lower limits from S09.  The black
  square (FGS) represents the best fit density for the entire S09
  survey, with $1\sigma$ error bars.  The estimates of the number of
  objects in the Classical Belt (CB), of Plutinos and Scattered Disk
  objects (SD) are plotted, as derived by \citet{1997Icar..127...13L},
  \citet{1997Icar..127....1M} and \citet{2008ApJ...687..714V}
  respectively, assuming each family is the unique precursor of
  JFCs.}\label{fig:uplim}
\end{figure}

We then process our implanted lightcurves as we previously did to
search for events, namely we filter, rank the lightcurves, and
evaluate the significance $\xi$ of each point in each lightcurve. We
then remove the false positives as described in Section~\ref{sec:fp}.
Our efficiency decreases rapidly with the KBO diameter: from nearly
33\% at $D=30~\mkm$ to $2\times10^{-5}$ at $D=0.5~\mkm$.  Note that at
$D=30~\mkm$ we ignore diffraction effects and the occultations are
modeled to suppress the flux completely for several consecutive
points, depending on the relative velocity, but our efficiency is
still significantly less than 100\%.  Some of our lightcurves are too
noisy to allow detections.

A set of synthetic events recovered by our pipeline in shown in
Figure~\ref{fig:fakeevt}, and the parameters of each plotted event are
given in Table~\ref{tab:fakeocc}.  Various parameters affect our
recovery efficiency. Our efficiency for the recovery of $D=3.0~\mkm$
occulting KBOs is plotted in Figure~\ref{fig:eff_vcrowd}.

The efficiency as a function of SNR is plotted in
Figure~\ref{fig:eff_vcrowd}a --a few targets at $\m{SNR}>100$ are left
out of the plot.  The behavior of our efficiency as a function of
crowdedness, where the crowdedness is defined as the number of targets
in that TAOS field brighter than $R_\m{USNO} = 13.5$, is plotted in
Figure~\ref{fig:eff_vcrowd}b. The efficiency is here plotted
multiplied by the number of targets in the field, to give a better
idea of the implication of this parameter for the event recovery. The
largest number of detections is achieved for more crowded fields.

The efficiency decreases with magnitude (Figure~\ref{fig:eff_vcrowd}c)
by about a factor of five between magnitude 9 and 13. The dominant
effect here is the decrease in SNR, though a competing effect occurs
since lower magnitudes are associated with larger angular sizes, and
our efficiency decreases with increasing angular size
(Figure~\ref{fig:angsz}). Furthermore there are many more dim than
bright stars in the sky: Figure~\ref{fig:eff_vcrowd}e shows the
efficiency as a function of $M_\m{TAOS}$ multiplied by the number of
TAOS targets at that magnitude. The highest number of detections
happen for stars with $M_\m{TAOS}\sim 12.5$.

Our efficiency as a function of the relative velocity of the KBO is
plotted in Figure~\ref{fig:eff_vcrowd}d. Observing at a pointing where
the relative velocity of the KBOs is higher boosts the event rate of
the survey. Our efficiency, however, is larger for smaller transiting
velocities, particularly for small KBOs for which the time-line of the
event is shorter than one of our data-points at
opposition. Ultimately, the effective sky coverage of our survey
depends linearly on both the efficiency and the velocity (see
section~\ref{sec:omega}).  The efficiency multiplied by the relative
velocity $v_\m{rel}$ is plotted against $v_\m{rel}$ in
Figure~\ref{fig:eff_vcrowd}f.  Pointing near opposition increases the
effective coverage of our survey, and thus it increases our event
rate, for 3~km KBOs. The survey strategy can be optimized at different
sizes taking into account the size dependent efficiency as well as the
expected KBO size distribution.

\subsection{Effective Sky Coverage and Upper Limits}\label{sec:omega}

We calculate the effective sky coverage of our survey, $\aeff$, as:
\begin{equation}
\aeff(D) = \frac{1}{w(D)}\sum_*
\frac{H(D,\theta_*)}{\Delta}~\frac{v_\mathrm{rel}}{\Delta}~E_*,
\end{equation} 
where $E_*$ is the exposure of the star target (the duration of the
lightcurve set), $\Delta$ the distance to the occulter, $H$ the cross
section of the event (Section~\ref{sec:analysis}), $w(D)$ the weight
factor for that diameter, i.e. the fraction of lightcurves implanted
with occultations by KBOs of diameter $D$, and the sum is carried out
only over the lightcurves where events are recovered. 

The effective coverage of our survey, which takes into account our
efficiency, is plotted in Figure~\ref{fig:Omegas}, for both the
dataset published in Z08 (empty squares) and for the current work
(filled squares). The solid line is a spline fit to the points.

\section{Model independent limits on the size distribution of KBOs}\label{otherocc}
\begin{figure}[bt]
\centerline{\includegraphics[width=0.5\textwidth]{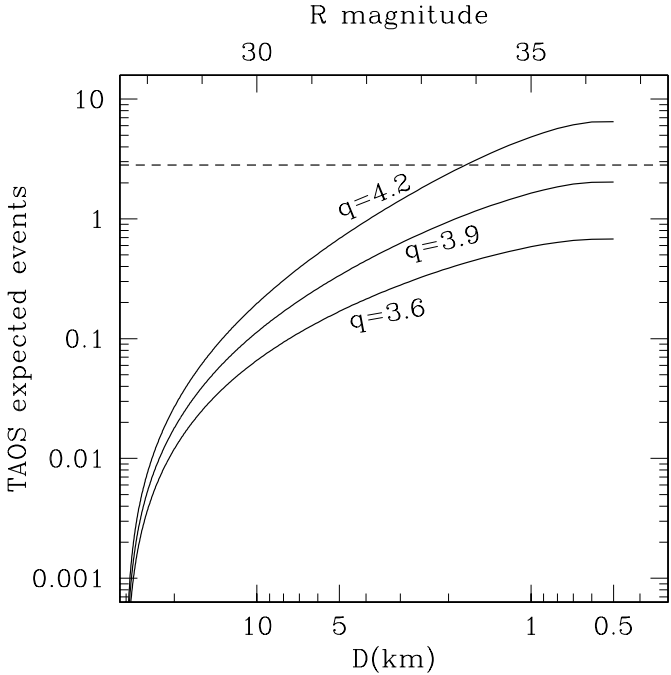}}
\caption{ Expected number of 
  events in 3.75 years of TAOS data from the size distribution presented in~S09. Slopes of $q~=~3.9$ (best fit to the HST/FGS data), $q=3.6$ and 4.2 ($\pm 1\sigma$ from the best fit) are used.\label{fig:fgs}   
}
\end{figure}

We can use $\aeff$ to calculate model-independent upper limits: at
each size for which our efficiency calculation was conducted we
calculate the number density of KBOs from the TAOS dataset as a
\emph{single-point} upper limit.  These limits are shown as filled
dots. We can then interpolate these upper limits with a spline fit,
obtaining the solid line in Figure~\ref{fig:uplim}. At each diameter
$D$ this represents the maximum surface density of KBOs of diameter
$\gtrsim D$. 

Figure~\ref{fig:uplim} shows model-independent upper limits reported
by the occultation surveys of \citet{2008AJ....135.1039B}, \citet{mmt}
and \citet{ps}.  The improvement is of nearly an order of magnitude at
700~m and over an order of magnitude at 1~km from the results of
\citet{mmt} and at 500~m from \citet{ps}. As discussed in
Sections~\ref{sec:intro} and \ref{sec:simul}, the TAOS sensitivity is
limited to relatively bright stars ($V\lesssim13.5$), and TAOS has a
relatively low efficiency at detecting $D<3~\mkm$ KBOs.  The large
star exposure of TAOS, however, provides a dataset that well
compensates for the lower sensitivity.

Limits from the recent HST/FGS occultation survey~\citep[hereinafter
  S09]{fgs} are included in Figure~\ref{fig:uplim}.  S09 reported the
detection of one possible event at a relatively high ecliptic
latitude.  The event is consistent with an occultation by a KBO with
$D=1.0$~km.  The dashed lines in Figure~\ref{fig:uplim} represent the
95\% c.l. upper and lower limits from S09. The TAOS upper limits are
more constraining than the corresponding limits from S09 for
$D\gtrsim0.6~\mkm$.  The black square (FGS) represents the best fit
density for the entire S09 survey, with $1\sigma$ error bars. Note
that this is not a model-independent estimate, as it is derived
assuming a straight power law distribution for small KBOs.  Note also
that TAOS has a greater sky coverage at $D\sim1~\mkm$, where the S09
event was detected, by a factor $\sim6$. This detection reported by
S09 is, however, not statistically inconsistent with our upper limit.

Figure~\ref{fig:fgs} shows the expected number of events for the TAOS
survey if one assumes the size distributions derived in S09. S09
estimated a cumulative size distribution $\Sigma_N(D>0.5~\mkm)=2.1
\times 10^7~\mathrm{deg}^{-2}$ from their detection and
non-detections, and fit a power-law to their data, anchored at
$\Sigma_N(D>90~\mkm)=5.4~\mathrm{deg}^{-2}$, to yield a differential
slope of $q~=~3.9\pm0.3$.  Figure~\ref{fig:fgs} shows the expected
number of events for the TAOS survey for these distributions.  We
calculate the number of events expected in our survey as:
\begin{displaymath}
N_\mathrm{exp} = \int\frac{dN}{dD}~\Omega_\mathrm{e}~ dD.
\end{displaymath}
With no detections our survey can rule out distributions that predict
$N_\mathrm{exp} \ge 3$ at the 95\% confidence level (c.l.).
The
best fit S09 model leads to an expected number of 2.1 detections in
our TAOS data.  This is a reasonable consistency.  
We can
rule out slopes steeper than $q~=~4.0$ for this model of the size
distribution at 95\% c.l.


Figure~\ref{fig:uplim} also shows the estimates of JFC progenitor
populations. Our constraints on these families are discussed in
Section\ref{sec:JFC}.

\section{Outer Solar System Collisional Models}\label{sec:models}

The collisional and dynamical evolution of the Solar System shaped the
size distribution of the Kuiper Belt.

The belt was originally
populated by very small dust grains, with small orbital eccentricities
($e \leq 0.01$) such as those we observe in circumstellar disks around
other stars~\citep{2007astro.ph..3383M}. Initially these small objects
merge and grow \citep{1999AJ....118.1101K}: 1~km KBOs in the Kuiper
belt are thus formed.
As their gravitational cross section grows larger than their geometric
cross-section, \emph{gravitational focusing} speeds up the growth rate
of the largest bodies.  This phase is referred to as runaway growth,
and objects as large as hundreds of kilometers can form.  One such
population, shaped primarily through agglomeration processes is
predicted by theory to have a power law distribution in diameter
$dN/dD \propto D^{-q_\mathrm{L}}$ with power $q_\m{L} \approx
4.5$~\citep{1999ApJ...526..465K}. Direct observations of large KBOs
confirm the power law behavior in this regime, the
\emph{gravitationally--dominated} region of the size spectrum, with a
best fit of $q_\m{L} = 4.8$ (\citealt{2009AJ....137...72F},
\citealt{2008arXiv0804.3392F} and reference therein).  The size
distribution of these large objects, for which gravity dominates the
internal strength, is remarkably insensitive to parameters such as
Neptune stirring or the internal tensile strength of the KBOs.

Meanwhile, very large objects in the planetary region of the Solar
System are also forming into planets, that are believed to undergo
significant migrations~\citep[and references
  therein]{2005Natur.435..459T}.  The orbits of the planetesimals are
then stirred up via gravitational interaction to velocities such that
further impacts will result in the disruption of the smaller objects:
this is the \emph{catastrophic collisions}
phase~\citep{1997Icar..125...50D, 1999AJ....118.1101K,
  2008ssbn.book..275M}; the time scale to reach this phase is
estimated to be between 10~Myr and 1~Gyr~\citep{2001AJ....121..538K}.
For very small objects (probably tens of meters and smaller) the
collisionally evolved population transitions to a regime where the KBO
binding energy is dominated by internal strength, rather than gravity.
Here the collisional cascade will generate a size distribution which
follows a power law with index $q_\m{S} =
3.5$~\citep{1969JGR....74.2531D, 1999AJ....118.1101K}, also in a
fashion that is largely independent on the details of the evolution of
the protoplanets. Note that the study of collisions between icy bodies
is still in its infancy, and future work in this field will permit
assessing the behavior of colliding small strength-less or loosely
bound particles~\citep{2008ssbn.book..195L}. Similarly, future work on
coupling collisional and dynamical evolution codes, recently pioneered
by ~\citet{2007Icar..188..468C}, should provide further insight in the
behavior of the size distribution.

 \begin{figure*}[t!]
\centerline{\includegraphics[width=0.5\textwidth]{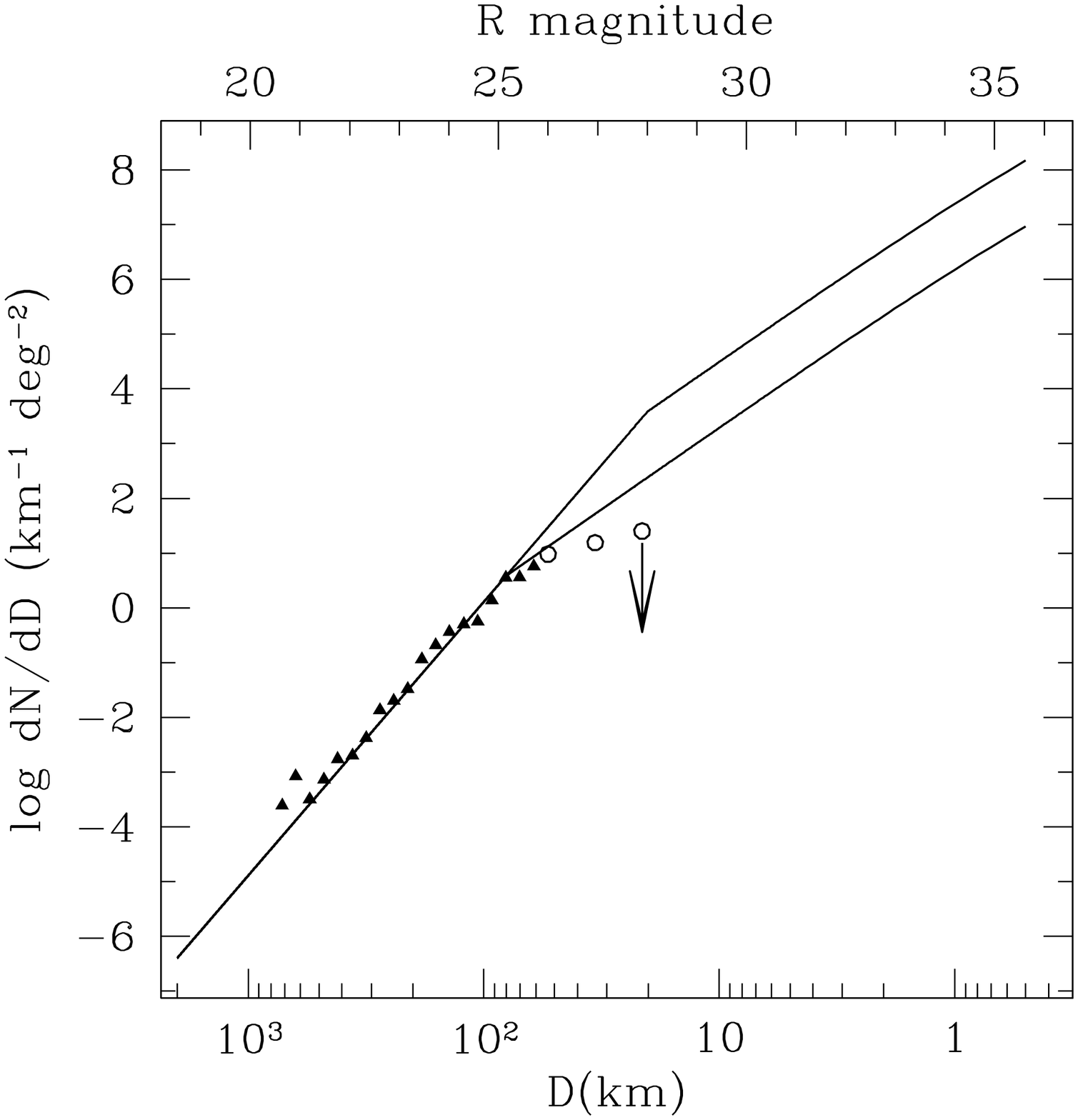}\includegraphics[width=0.5\textwidth]{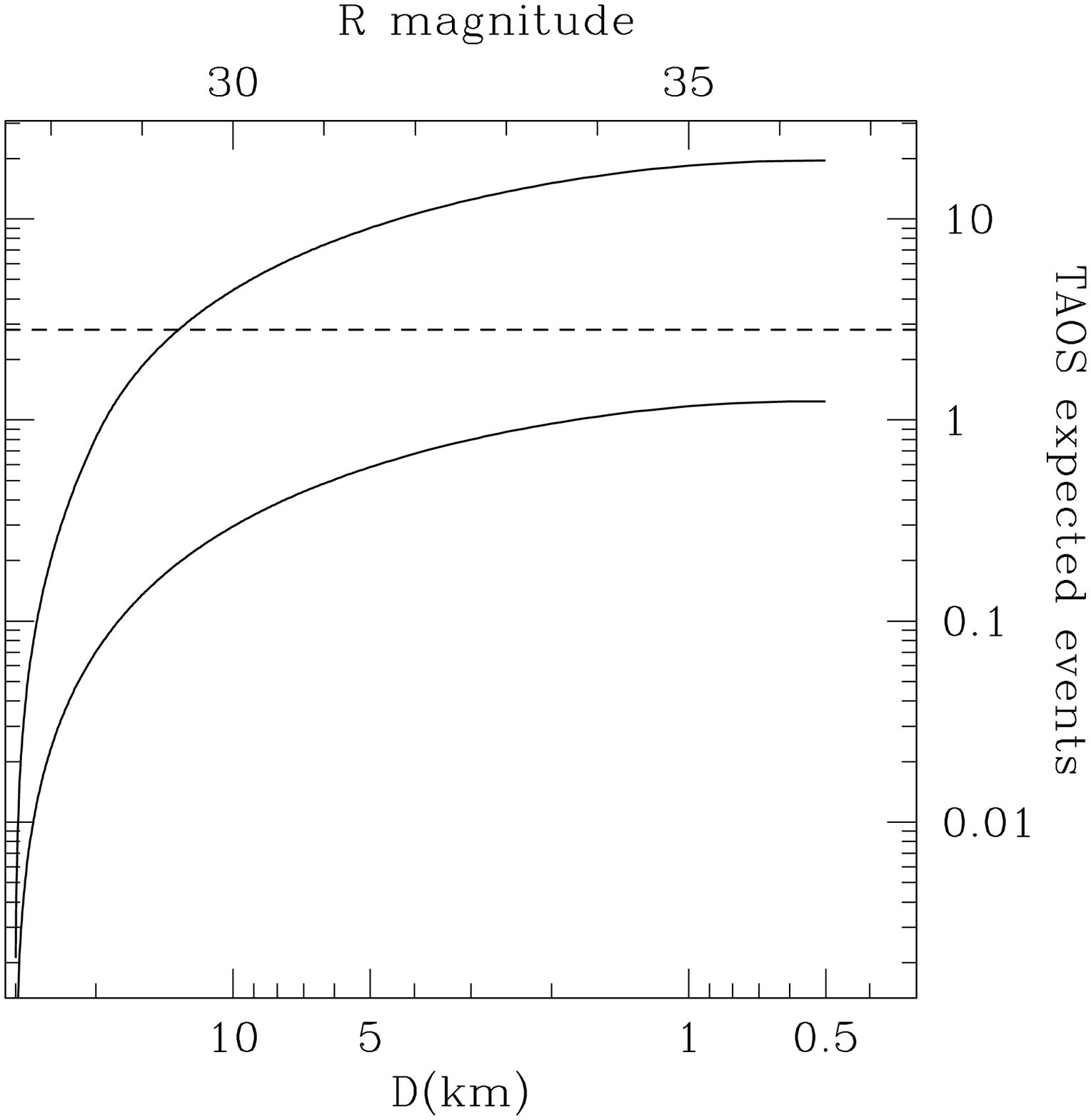}}\caption{PS05 model and expected event rate for TAOS. \emph{Left.}
  Triangles are the data from ~\citet{2009AJ....137...72F}. Empty
  circles are the data from \citet{2004AJ....128.1364B}. We use these
  data from direct imaging surveys to set the location of the large
  end size distribution. The model is parametrized with two slopes,
  $q_\m{L} = 5$ and $q_\m{I} = 3$. The positions of the break are plotted at
  $\Db = $80, and 20~km: limiting values for the PS05 models. \emph{Right.} Expected yield of
  events in 3.75 years of TAOS data. The horizontal dashed line
  represents the highest number of events allowed given no detections
  in the TAOS data ($\leq3$ events at the 95\% c.l.). Any
  model above this limit is ruled out by our
  survey.}\label{fig:pansari}
\end{figure*}
We will refer to the region in between these two regimes as the
\emph{intermediate} region.  The extent of, and behavior of the size
distribution in
the intermediate region are instead very sensitive to
the formation and evolution parameters, and observational information
on this region can be compared to evolution models. In general, for a
weaker KBO population the transition to the $q_\m{S} = 3.5$ power law
behavior will occur earlier, reducing the size of the transition
region. A strong KBO population will display an extended intermediate
region, generally showing here oscillations around a mean power law of
slope which also depends on the details of the population and its
evolution. In this section, we present four models, from literature, 
of
the KBO size distribution $dN/dD$, and a simple parametric model. On
the basis of these models we calculate the number $N_\m{exp}$ of
events expected to be detected by the TAOS survey. Since no events
were found in our survey, any model which predicts $N_\mathrm{exp} \ge
3$ is ruled out by TAOS at the 95\% c.l.

We caution the reader that the predictions presented here depend on
both the shape of the size distribution and on scaling parameters. All
of our models are scaled so that the differential size distribution is
consistent with the direct observations at $D~=~200~\mkm$, where the
direct surveys are most constraining: 
\begin{equation}
dN/dD(D=200~\mkm)\sim 0.04 ~\m{deg}^{-2}\mkm^{-1}
\end{equation}. 
For each model we will point
out the cumulative number of KBOs predicted at $D=100~\mkm$,
$\Sigma_N(D>100~\mkm$), with this scaling choice. Strengths and
weaknesses of all of these models are also discussed in
\citet{fraser09}.

\subsection{Pan \& Sari (2005)}

\citet[hereinafter PS05]{2005Icar..173..342P} derived a fully
analytical model for the size distribution of KBOs by assuming the
population is in a steady state and the mass is constant through the
collisional processes. They assume for most of their model calculation
that the internal strength of the objects is negligible (gravity
dominated objects). This assumption is motivated by studies of comets
and asteroids (PS05 and references therein).  The transition
to the fully strength-dominated regime, where the size distribution
follows \citet{1969JGR....74.2531D} with a power law with slope
$q_\m{S}=3.5$, occurs at $D\le300~\m{m}$. This region of the size
spectrum is entirely below the sensitivity of TAOS. 

PS05 derive an analytical double power law size distribution for
objects $D\geq300~\m{m}$:
\begin{eqnarray}
dN/dD&&\propto d^{-q_\m{L}}~~ for~~ D ~>~D_\m{b},\nonumber\\
dN/dD&&\propto d^{-q_\m{I}}~~ for~~ D~ <~D_\m{b}. 
\end{eqnarray}
This model is shown in Figure~\ref{fig:pansari}, left.  The slope $q$
has value $q_\m{L} = 5$ for large objects and $q_\m{I} = 3$ for
objects in the intermediate region.  PS05 are thus able to calculate
self-consistently the location of the break in the power $D_\m{b}$, which
represents the size of the largest KBOs that experienced catastrophic
collisions, as a function of time. The location of the break moves
toward larger objects as the size distribution evolves.
\begin{figure}[t!]
\centerline{\includegraphics[width=0.5\textwidth]{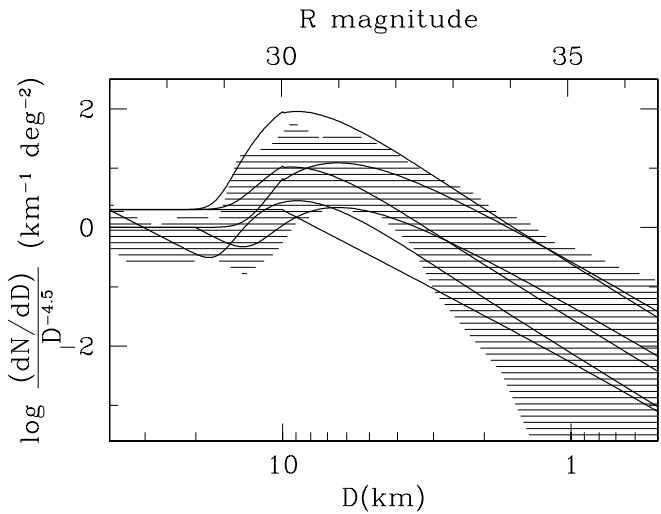}}
\caption{Our modeling of the KB04 results.  The range of results of
  the simulations of KB04 is represented by the shaded region (see Figure 10 and 11 in KB04). A few
  of our models are plotted as solid lines. All models are normalized
  by the slope of the large KBO size distribution
  $D^{-4.5}$.}\label{fig:kb0}
\end{figure}

Model distributions with break points $\Db~=~80$, and
20~km, limiting values for PS05, 
and $\Sigma_N(D \geq 100~\mkm) \sim 32~\m{deg}^{-2}$
are shown in
Figure~\ref{fig:pansari}. 
The corresponding predicted number of events for the TAOS data
analyzed in this paper is plotted in the right panel\footnote{ Note
  that a simple extension of the large-size power law distribution,
  which represents the naive expectation before
  \cite{2004AJ....128.1364B} appeared, would lead to thousands of
  detections. This was the initial design target of the TAOS
  project.}.

structure of the size distribution in the intermediate region is
however generally more complicated. Between $\Db$ and the second
break, which is the region that TAOS can probe, a realistic size
distribution is expected to have an oscillatory behavior which in the
PS05 models preserves an average slope of $q_\m{I}=-3$ (PS05). These
waves however, as they are described in PS05, are small in amplitude
and affect our prediction to less than 1\%.
On the basis of our no-detection result, break diameters of $\Db <
51.3~\mkm$ are excluded at 95\% c.l.  This is consistent with the data
from direct observations and with the authors' interpretation of the
model: in absence of stirring by Neptune, the location of the break is
consistent with a KBO population comprised of objects with little
internal strength. Note that direct surveys also suggest that for one
such distribution the location of the break should be at a large
diameters (see~\citet{2009AJ....137...72F} and
Figure~\ref{fig:pansari}, left).
\begin{figure*}
\centerline{\includegraphics[width=0.5\textwidth]{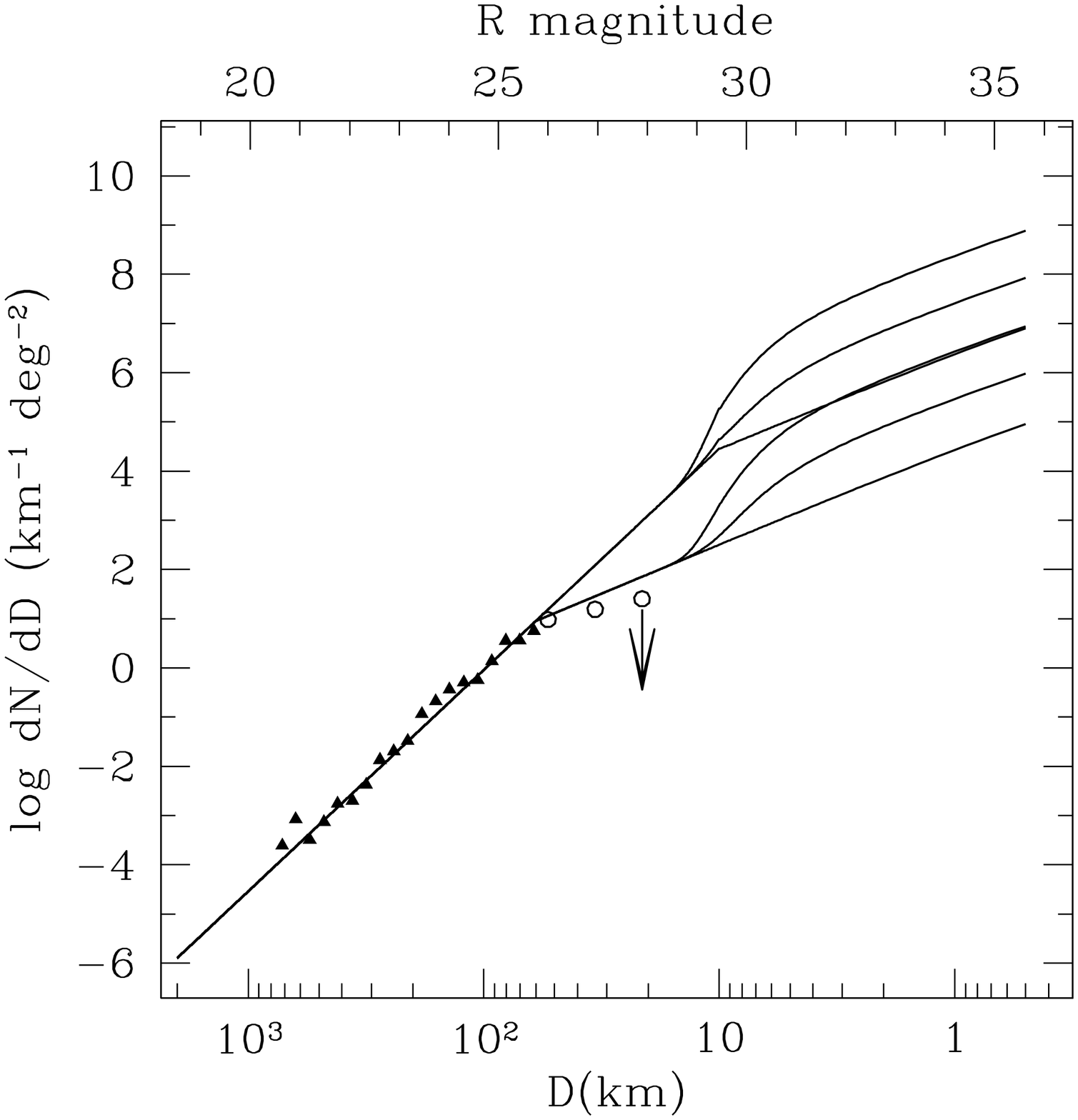}\includegraphics[width=0.5\textwidth]{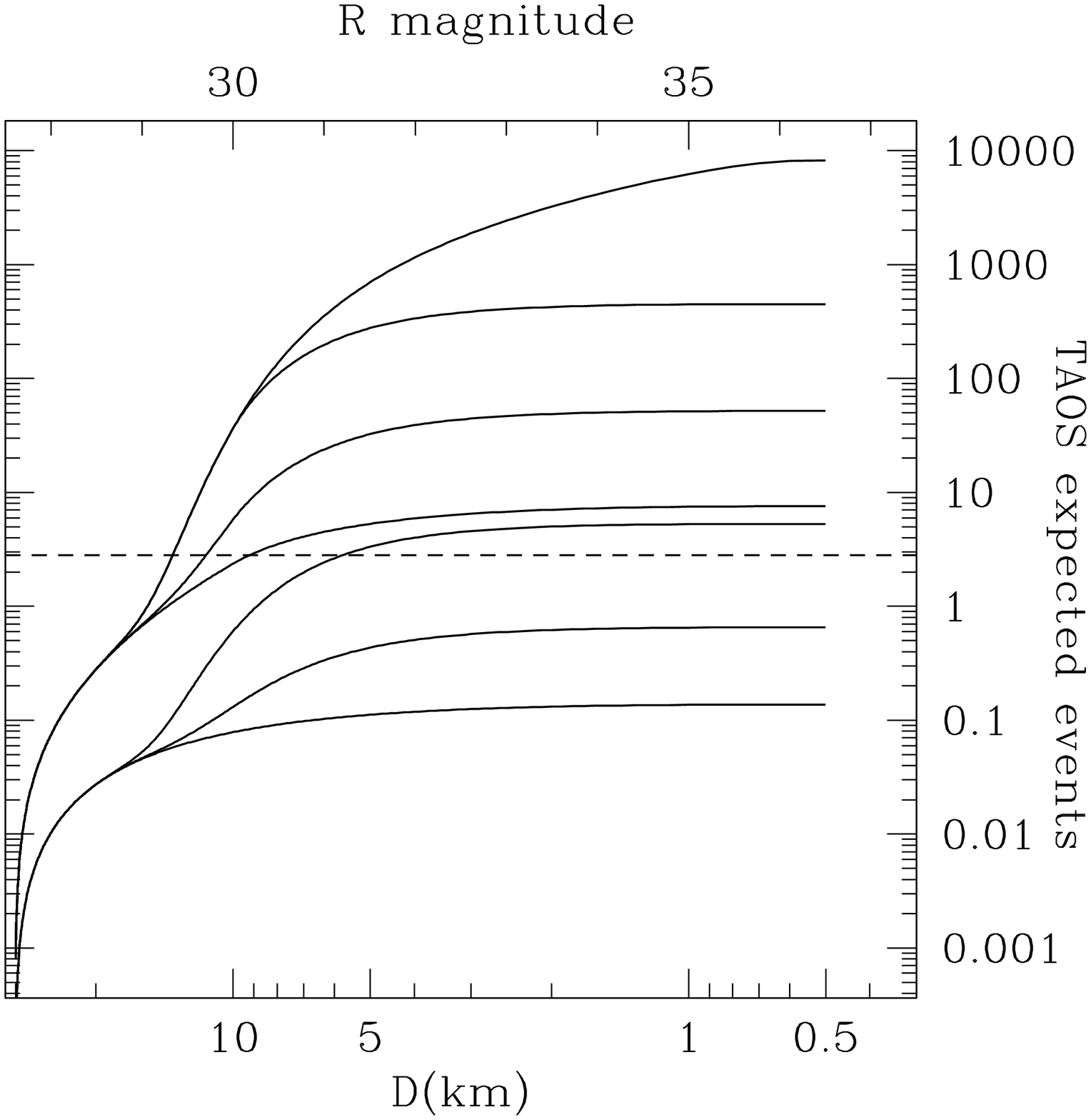}}

\caption{KB04 models. The left panel shows the differential size
  distribution, parametrized as two slopes ($q_\m{L}=4.5$ and
  $q_\m{I}=2$) and on the right side are the corresponding event
  rates for the TAOS survey. Symbols are the same as in
  Figure~\ref{fig:pansari}. For limiting break positions of $D_\m{b}=60$ and 10~km the distribution is plotted for a model with no Gaussian excess, and models with Gaussian excess of intensity
  $I_\m{ex}=10$ and $I_\m{exp}=100$, centered at  $\mu_\m{ex} = 5.5$~km.
  }\label{fig:kenbrom1}
\end{figure*}

\subsection{Kenyon \& Bromley (2004)}\label {sec:kb}

\citet[hereinafter KB04]{2004AJ....128.1916K} developed numerical
models of the collisional evolution of KBOs, including the effects of
the internal strength and
gravitational binding of KBOs, as well as the initial mass in the belt
and a model of stirring by Neptune: if in its migration Neptune
reaches its current location early its stirring effect will influence
the shaping of the size distribution of the Kuiper Belt.  The
disruption energy, defined as the energy necessary to remove 50\% of
the combined mass of the colliding bodies, is modeled as
$Q_d~=~Q_br^{\beta_b}+\rho~Q_gr^{\beta_g}$, where $Q_b$ and $\beta_b$
describe the internal binding energy and $Q_g$ and $\beta_g$ the
gravitational energy, with $\rho$ the density and $r$ the radius of
the object. In the KB04 simulations $Q_b$ is varied between $10$ and
$10^8~\mathrm{erg~g^{-1}}$, $\rho~Q_g$ between $10^{-4}$ and $10^4$
and $\beta_g$ between 0.5 and 2.0, while $\beta_b$ is set to 0, as
this parameter has little effect on the simulation results.

All KB04 models generally agree in the shape and slope of the size
distribution for large KBOs ($D=80~\mkm$ and larger) generating a
\emph{cumulative} size distribution which follows
$N(D\gtrsim80~\mkm)\propto D^{3.5}$, equivalent to a power law
differential size distribution with power $q_\m{L} \approx 4.5$, in
good agreement with data and theoretical predictions for the evolution
in the gravitationally-dominated regime. The models display a variety
of behaviors for smaller objects. In all simulations a small dip in
the 10-40~km region is predicted (cf Figures 10 and 11 in KB04). This
is roughly consistent with the results of \citet{2004AJ....128.1364B}:
here collisions destroy weak KBOs and models with Neptune stirring or
weakly bound KBOs produce a more significant dip. This feature is
followed by an excess with respect to the nominal power law for 2-15
km KBOs. The amplitude of the excess varies
substantially, 
between a factor
of a few and a factor of a few tens, depending on the internal
strength of the objects and on the 
details of the effect of Neptune stirring.
The
size distribution remains sensitive to the details of the models down
to about $D=50~\m{m}$, where once again a power law behavior begins,
with power $3<q<5$ in the strength-dominated regime.

Figures~\ref{fig:kb0} shows our 
representation 
of the models in KB04,
as they are presented in Figure 10 and 11 of KB04. 
The size distribution is here shown scaled by the
slope of the large size region ($D^{-4.5}$). 
The shaded region
represents the 
range of the KB04 models and the plotted
lines are some of our models, covering a large region of this
model space. 
Focusing on the region
near the transition between primordial and collisionally evolved
population, an excess near 5~km and a depletion near 25~km are both
visible and well represented in our models.  

We model the large size distribution as a power law with
$q_\m{L}~=~4.5$ and the intermediate region with $q_\m{I} = 2$.  We
model the excess near $D~=~5~\mkm$ with a Gaussian, so that:
\begin{equation}
\frac{dN_\m{ex}}{dD} ~=~\frac{dN}{dD}~\left(1~+~I_\m{ex}\exp
-\frac{(D-\mu_\m{ex})^2}{2.0~\sigma_e}\right).
\end{equation}
We fix the width of the Gaussian excess to $\sigma_e = 3.5~\mkm$, we set the location to $\mu_\m{ex} = 5.5~\mkm$ or $\mu_\m{ex} = 1.6~\mkm$. The intensity of the excess is determined by $I_\m{ex}$; in order to fairly represent all results from the KB04 simulations we consider models with an excess of $I_\m{ex}~=~0,~10$ and ~$100$. Note that as the break diameter moves towards large sizes the models naturally simulate the small dip near 20~km (Figures~\ref{fig:kb0}). Our results are not sensitive to the presence of this small depletion. With our scaling we obtain a cumulative surface density of KBOs $N(D \geq 100~\mkm)\sim
25~\m{deg}^{-2}$. 

Figure~\ref{fig:kenbrom1} shows the models with $\mu_\m{ex} =
5.5~\mkm$ (left) and the corresponding expected number of events for
the TAOS dataset (right) for the limiting break positions
$D~=~60~\mkm$ and $D~=~10~\mkm$ for $I_\m{ex} = 0,~10,~\m{and}~100$.
For a given excess intensity, we can constrain the location of the
break: in absence of the excess break diameters smaller than $\Db =
16.5~\mkm$ are ruled out; break diameters $\Db < 32.6~\mkm$ are ruled
out for $I_\m{ex} = 10$ and $\Db < 75.3~\mkm$ are ruled out for an
$I_\m{ex} = 100$ excess. When moving the excess towards smaller sizes
($\mu_\m{ex} = 1.6~\mkm$), break locations $\Db < 28.3~\mkm$ are ruled
out for $I_\m{ex} = 10$, and $\Db < 63.3~\mkm$ for $I_\m{ex} = 100$.

Since the KB04 simulations show that models with weaker KBOs
and Neptune stirring produce a location of the break at smaller sizes, 
our result strongly favors models that incorporate the
effects of Neptune stirring and weaker KBOs, where the bulk strength
$Q_b \lesssim 10^3~\m{erg}~\m{g}^{-1}$

\begin{figure*}[t!]
\centerline{\includegraphics[width=0.5\textwidth]{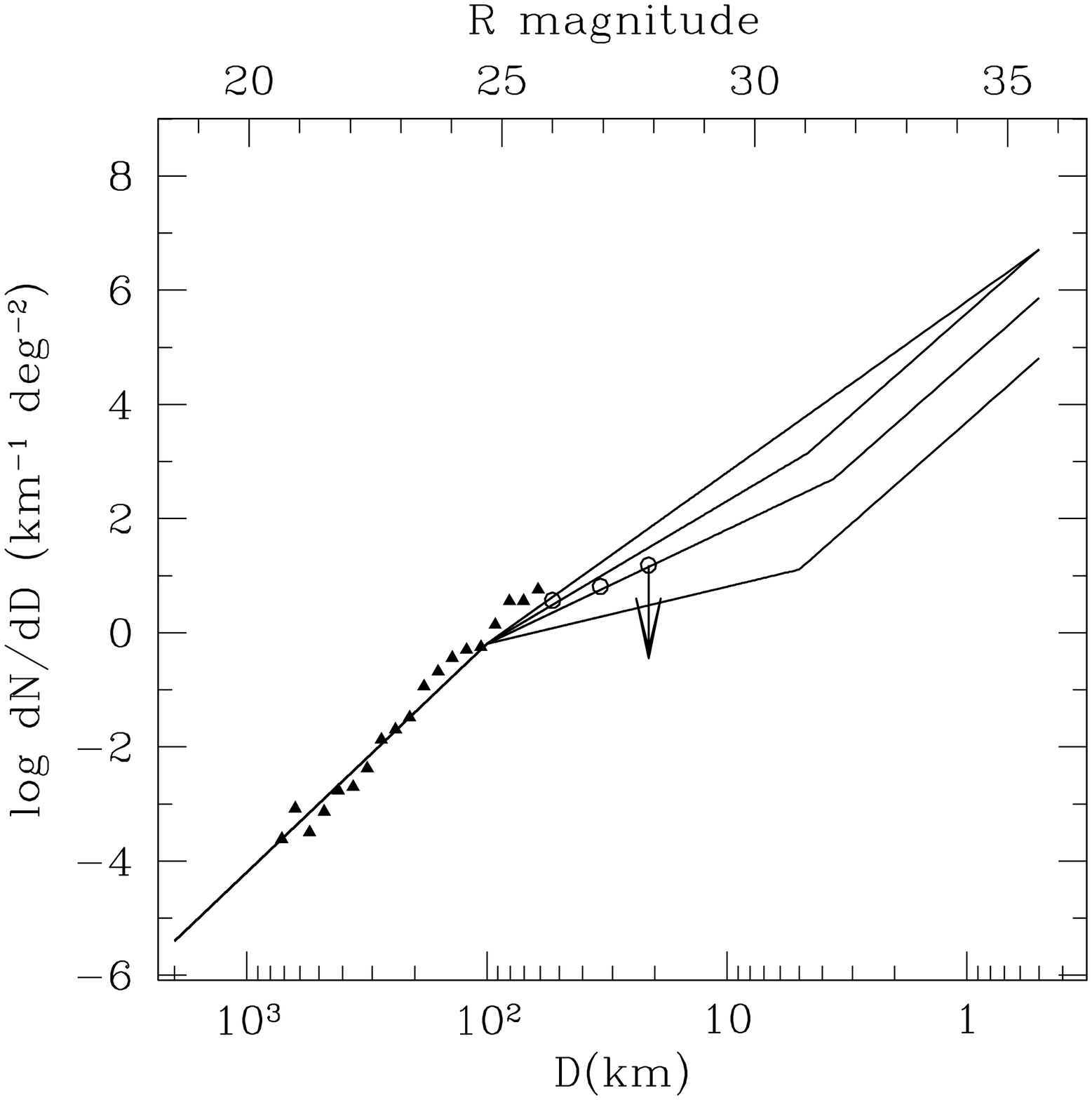}\includegraphics[width=0.5\textwidth]{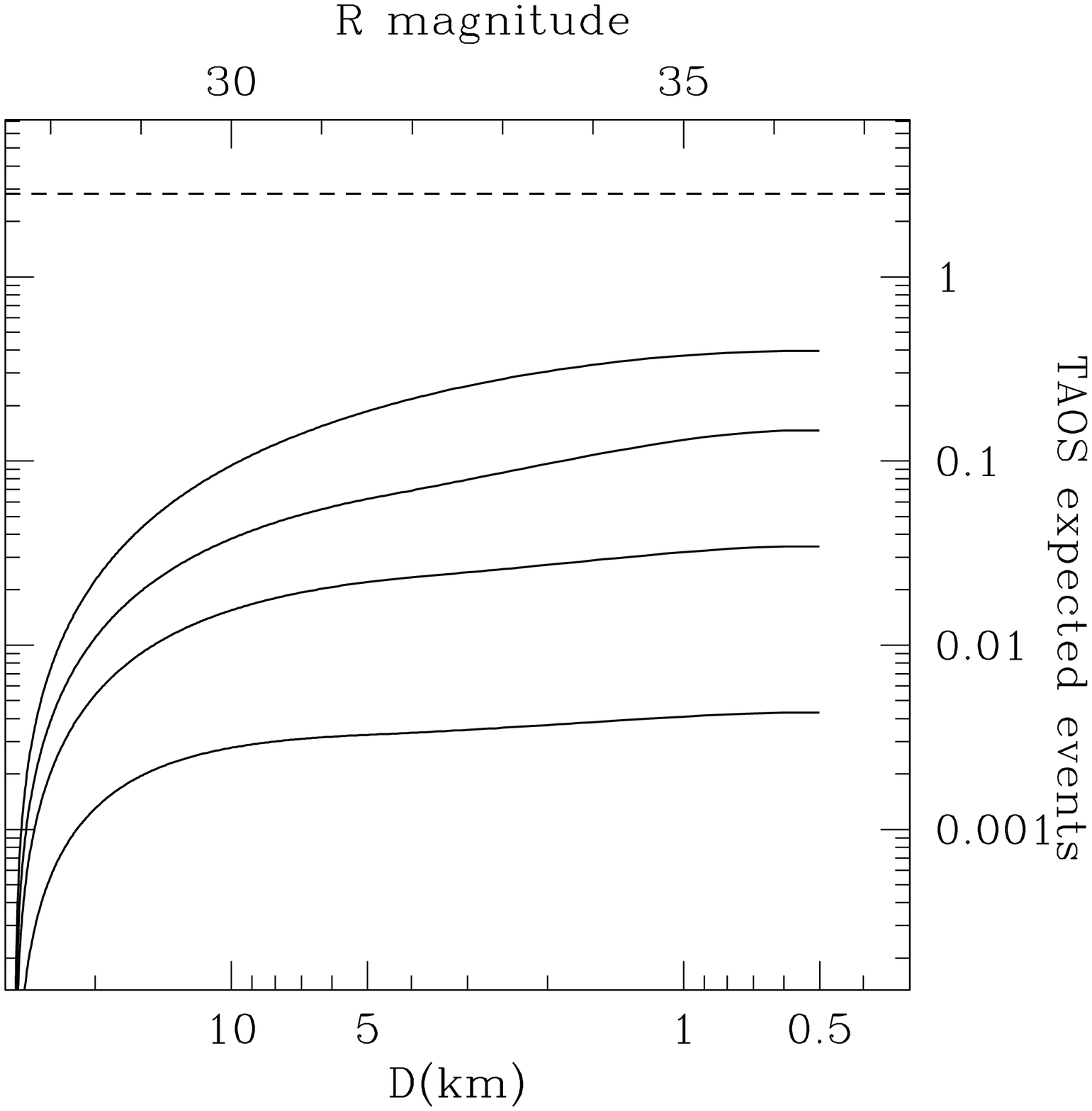}}

\caption{Our rendering of BCB09 models: the models differ in the
  prescription for fragmentation and all models are parametrized as a
  series of three slopes. Symbols are the same as in
  Figure~\ref{fig:pansari}. The first break point is fixed at
  100~km. The slope for the smallest size objects is set to 3.7. The
  location of he second break and intermediate slope are 5~km and 1.0,
  3.6~km and 2.0, 4.6~km and 2.5 and 0.36~km and 3.0.  Right:
  corresponding number of events in our TAOS
  survey. }\label{fig:benavidez}
\end{figure*}

\begin{figure*}[t!]
\centerline{\includegraphics[width=0.5\textwidth]{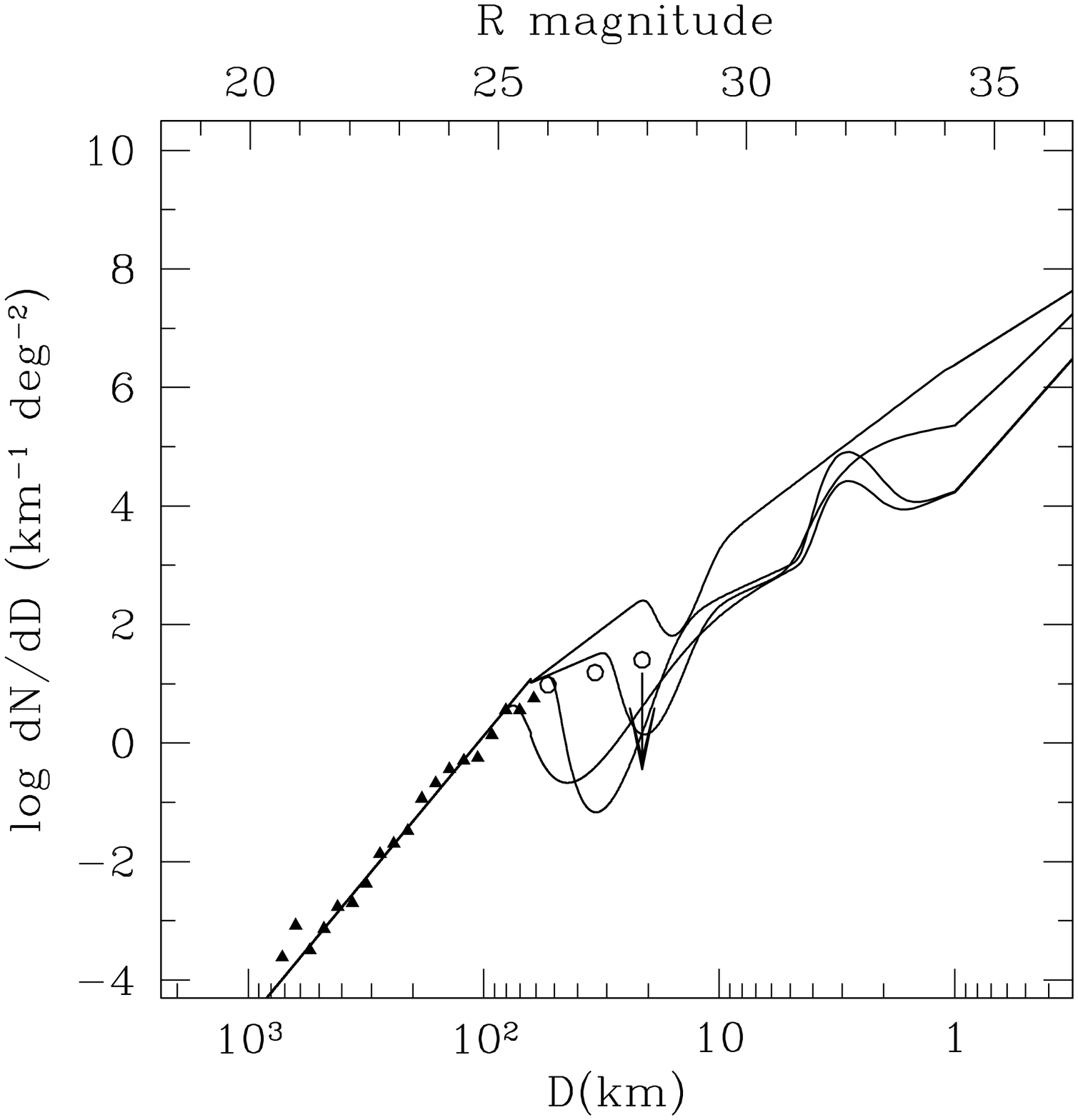}\includegraphics[width=0.5\textwidth]{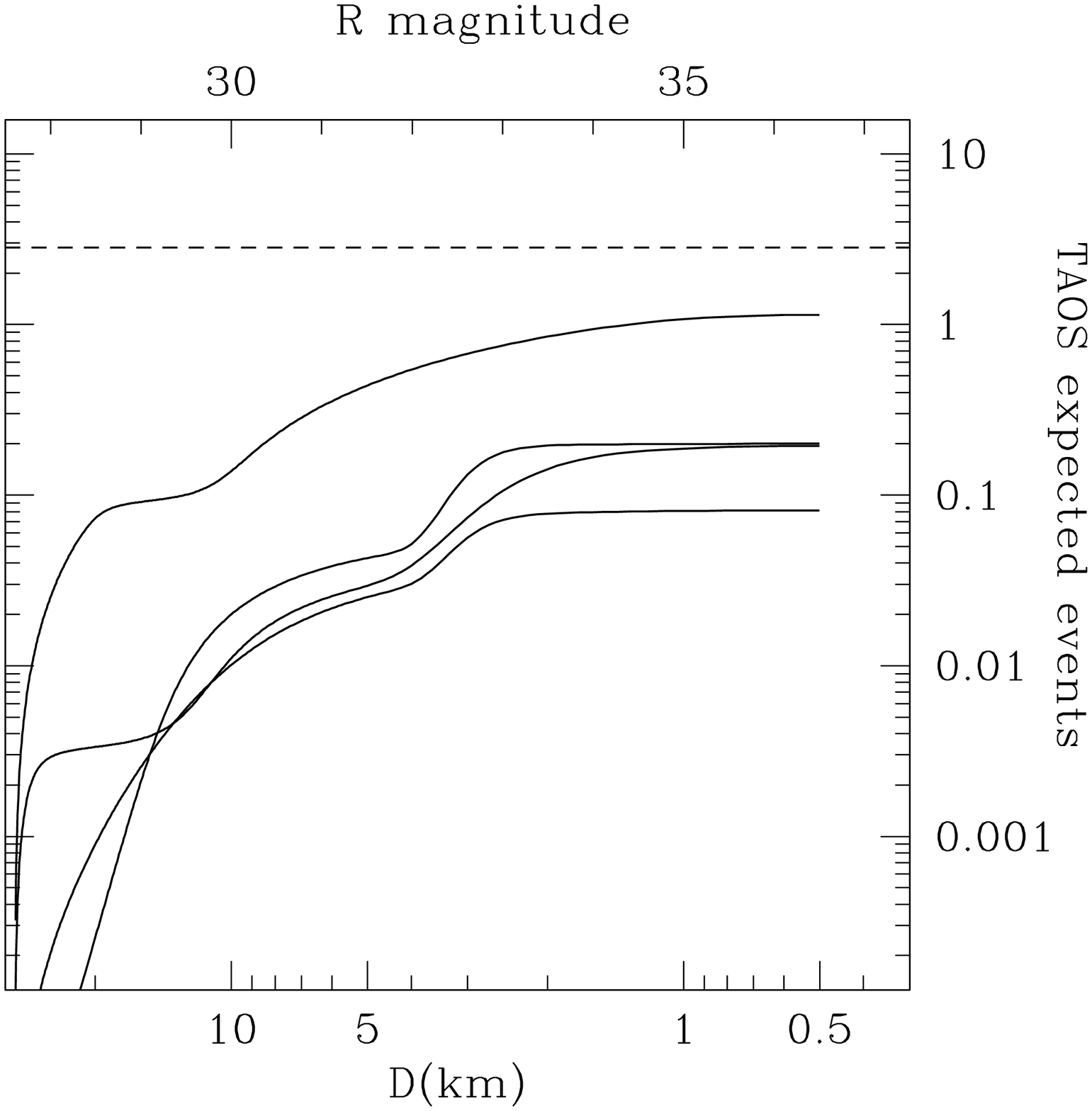}}

\caption{Left: Our modeling of the F09 models. The models are
  reproduced on the basis of Figure~2 in F09, and divots and excesses
  are generated with Gaussians. Symbols are the same as in
  Figure~\ref{fig:pansari}. The slope on the large hand side
  ($D~\gtrsim ~65~\mkm$) is $q_\m{L}=4.8$. Right: corresponding event
  rates for our survey.}\label{fig:f09}
\end{figure*}

\subsection{Benavidez \& Campo Bagatin (2009)}\label{sec:bena}
Recent simulations by \citet[hereinafter BCB09]{2009P&SS...57..201B}
divide the Kuiper Belt into three dynamical families -- the CB, the
Plutinos (Resonant Population) and the SD -- and follow the
collisional evolution of each, while taking into account the physics
of the fragmentation of icy and rocky bodies at the typical relative
velocities of KBOs. This suite of models ignores the effects of
Neptune stirring. The models incorporate four scaling laws for
fragmentation: a simple scaling driven by gravitational
self-compression, two scaling laws which include both self-compression
and the effects of strain-rate, as described by
~\cite{1982Icar...52..409F}, with different diameter dependency
($D^{-0.25}$ and $D^{-0.5}$) and one that follows the modeling of
~\cite{1999Icar..142....5B} for icy bodies. They then vary the
material strength of the objects between $10^5$ and $10^7
\m{erg}~\m{cm}^{-3}$.

In BCB09 the size distribution of large objects is set to a power law
with slope $q_\m{L} ~\approx 4.0$. This is set as an
initial condition to their simulations, and since the large objects are
not undergoing many collisions, this shape is preserved. A first
break is seen around $100~\mkm$.  The size
distribution then departs from power law behavior for all parameter
choices, with two to four orders of magnitudes fewer KBOs in the
$\sim1~\mkm$ region than the nominal power law would predict. The size
distribution then follows again a power law, with slope $q_\m{S} =
3.5$, in the strength-dominated regime. The BCB09 models are shown in
Figure~\ref{fig:benavidez}, left. We simplified the size distribution
behavior in the intermediate region with a single power law, though
typically the behavior is oscillatory. Our scaling leads to 
$N(D \geq 100~\mkm)\sim 21~\m{deg}^{-2}$ for the BCB09 models. 

These models are all allowed by the TAOS data, partly because of the
location of the initial break at $D_\m{b} = 100$ and the slope of the
large end size distribution. The break location is at the large end of
what is allowed by direct observations (indeed outside of
the~\citet{2008Icar..198..452F} allowed range of
$D_\m{b}\in[50,95~\mkm]$ with assumption of a 6\% albedo). The
location of this first break does not evolve in the BCB09 simulations
from what is set as an initial condition. Similarly, the choice of a
slope at the large end of the size distribution of $q_\m{L}=4.0$ is
slightly shallower than the current best fit value
from~\citet{2009AJ....137...72F} and \citet{2008arXiv0804.3392F}. As
the parameters relative to the large size end of the distribution are
more firmly pinned down by direct observations, these models and
future occultation data may place stronger constraints on the details
of the shape of the size distribution below the first break and thus
the details of the fragmentation mechanisms, providing information on
the internal structure of the KBOs.
  
\subsection{Fraser (2009)}

\citet[hereinafter F09]{fraser09} considers the collisional evolution
of the Kuiper Belt size distribution \emph{after} the epoch of
accretion. Starting with initial conditions that reproduce the
observed large end size distribution ($q_\m{L}=4.8$), which will not
further evolve in the collisional simulations, and $D_\m{b1}=2~\mkm$,
the population is collisionally evolved over the age of the Solar
System. A depletion, or \emph{divot}, forms at $D~\sim10\times
D_\m{b1}$, or $D\sim20~\mkm$. The size of this depletion changes when
changing parameters relative to the internal strength of the KBOs or
the impact velocity, as well as choices for the initial intermediate
slope. An excess is also evident at smaller sizes: $D\sim4~\mkm$.
We model the F09 size distributions as they appear in Figure~2 of
F09. Our size distributions are created as a sequence of 3 slopes:
$q_\m{L}=4.8,~ 1\leq q_\m{I} \leq 3$, and $2.5\leq q_\m{s}\leq 4.5$,
and we model the divots and excesses with Gaussians
(Figures~\ref{fig:f09}, left). Here we scale the differential
size distribution so that $dN/dD(D=200~\mkm)\sim
0.047~\m{deg}^{-1}\mkm^{-1}$, to reproduce the number of objects at
$D=50~\mkm$\footnote{$N(D=50~\mkm)\sim15$, Fraser, private
  communication.}.
Our survey cannot constrain these models: the pronounced divot at
$D~\sim~20~\mkm$ causes a very low event rate in our survey
(Figure~\ref{fig:f09}, right).


\subsection{Generic 3--Regime Model: Constraints on the Intermediate  Region of the Size Spectrum}\label{sec:3par}

A simple, generic 3--regime model allows us to describe
separately the primordial region, the intermediate region and the
fully collisional, strength-dominated regime.  Knowing that both the
large (gravitationally-dominated) and the small (strength-dominated)
regions of the size spectrum are relatively insensitive to the details of
the internal structure and evolution of the Kuiper Belt, it is the
intermediate region that contains the most information about the
physical details of the KBOs. We grossly simplify the
expected structure in the intermediate regime and describe it with a
power law. The parameters of one such models would then be three slopes
$q_\m{L}, ~q_\m{I}$, and $q_\m{S}$, and two break locations
$~D_\m{b1}$ and $D_\m{b2}$.

\begin{figure*}[t!]
\centerline{\includegraphics[width=0.5\textwidth]{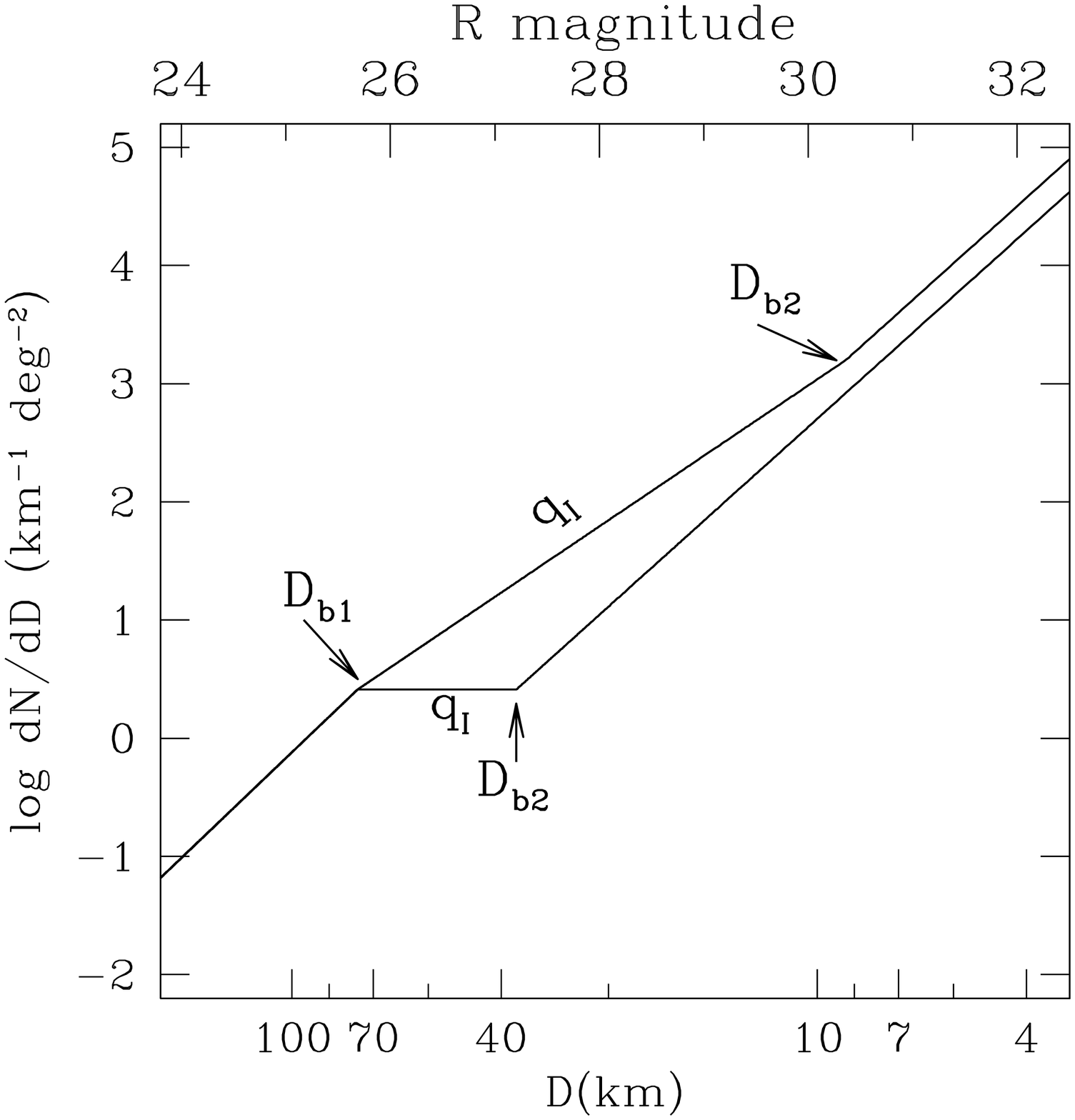}\includegraphics[width=0.5\textwidth]{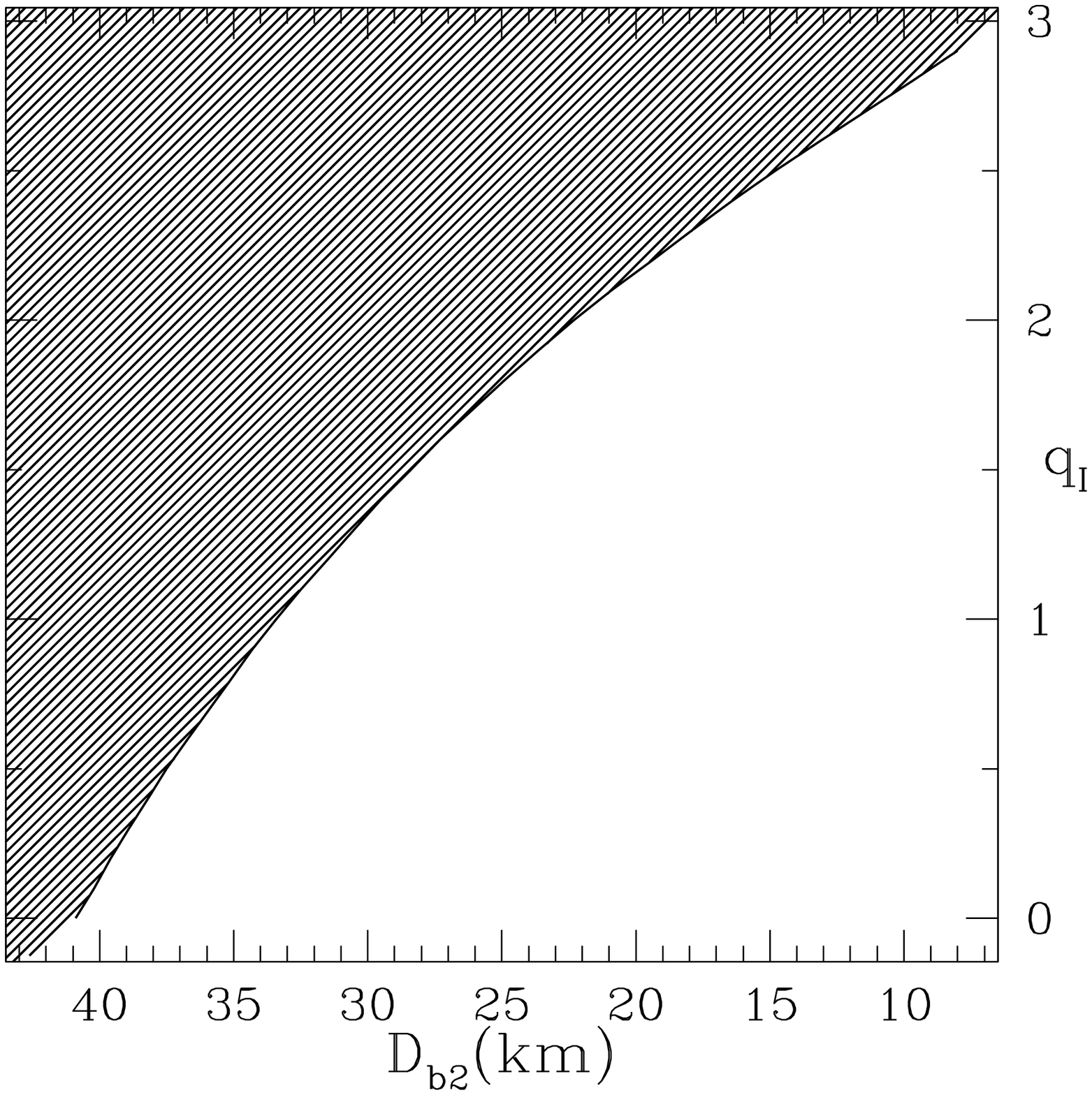}}
\caption{Three-slope model described in Section ~\ref{sec:3par} (left)
  for the minimum second break location $D_\m{b2}$ allowed by slopes
  $q_I = 0$ and 3. Slope--break location phase space for the
  three-slopes model (right). The shaded region is excluded by
  our data.}\label{fig:3par_exc}
\end{figure*}

 We set the slope of the large end of the size spectrum and the
 location of the first break to the best fit to the data from direct
 observations: $q_\m{L} = 4.8$ and $D_\m{b1} = 75~\mkm$. We model the
 intermediate region as a plateau, or with a shallow slope: $q_\m{I}
 \in [0, 3]$, and the small size end of the spectrum as a power law
 with slope $3.5$ for the strength-dominated, collisionally evolved
 population. Figure~\ref{fig:3par_exc} shows our 3-regime model (left)
 and the limit we can set to the intermediate slope, $q_\m{I}$, and
 second break location, $D_\m{b2}$, phase space (right). Any pair of
 values $q_\m{I}$ and $D_\m{b2}$ that fall above the solid line are
 excluded.  

\section{The Jupiter Family Comets Progenitor Population}\label{sec:JFC}
The Jupiter Family Comets (JFCs) are believed to originate in the
Outer Solar System. In this scenario the giant planets generate
gravitational perturbations that affect the orbits of the Outer Solar
System bodies, injecting them into the planetary region, where they
are captured by Jupiter. The orbital inclination of the JFCs suggests
that their precursor population has a disk-like distribution, favoring
thus the Kuiper Belt over the Oort Cloud as a reservoir~\citep[ and
  references therein]{2008ApJ...687..714V}. The Classical Belt (CB),
the Plutinos and the Scattered Disk (SD) have been considered as
precursors in various studies. The dynamical characteristics of each
population determines the efficiency of the injection process and the
number of objects in each progenitor family can thus be derived on the
basis of the density of JFCs, which is observationally constrained
(see~\citealt[and references
  therein]{2006Icar..182..527T}). Furthermore the size distribution of
JFCs should reflect the size distribution of the progenitor
population. \citet{2004AJ....128.1364B} and \citet{mmt} have pointed
out that a better determination of the size distribution of the Kuiper
Belt would help understand the origin of the JFCs.

We assume the JFC precursors are in the size range $1-10~\mkm$ as this
is observed to be the typical size of JFCs
\citep{2008ssbn.book..397L}.  We considered the estimates on the KBO
populations (CB and Plutinos) and SD derived from dynamical
simulations under the assumption that each population is the unique
progenitor of the JFCs, and we compared them to the upper limits derived
from our survey (Figure~\ref{fig:uplim}).  We use the estimate of
\citet{1997Icar..127...13L} for a population of cometary precursors
entirely in the CB, and that of \citet{1997Icar..127....1M} for
Plutino progenitors. These are converted into a surface density by
assuming for each population a projected sky area of
$10^4~\mathrm{deg}^2$, as was done by \citet{2004AJ....128.1364B}.  We
consider the results of~\citet{2008ApJ...687..714V} for a progenitor
population in the SD .  We calculate the minimum surface density of SD
objects expected in the region of sky typically observed by the TAOS
survey. For this we use information on the fraction of time the
objects spend between 30 and 50 AU and within $3\degree$ of the
ecliptic plane as provided by ~\citet{2008ApJ...687..714V}. These
estimates on the number of objects are shown in Figure~\ref{fig:uplim}
as horizontal lines.

Our results rule out a precursor family composed uniquely of CB
objects with $D\gtrsim3~\mkm$.  Occultation surveys are the only surveys
that at present can probe this region of the size spectrum, and our
preliminary result shows that future occultation surveys will be able
to derive useful constraints on the origin of the JFCs.

\section{Conclusions}\label{sec:conc}

We presented an analysis of 3.75 years of TAOS data, comprising
$5\times10^5$ star--hours of lightcurves sampled at 4 or 5~Hz observed
with three telescopes simultaneously. We searched for occultations of
our
target stars 
by KBOs in order to constrain the size distribution
of KBOs, particularly in the 500~m to 10~km region, which is currently
out of reach of direct observation surveys.  
More than 90\% of the TAOS data is collected within $5\degree$ of
ecliptic latitude in order to maximize the occultation rate by objects
in the Kuiper Belt. Occultations near opposition lead to a higher
event rate for TAOS, even after taking into account the increased
recovery efficiency for small objects where the angle from opposition
is larger and the relative velocity of the KBOs is lower.

We found no occultation events in our data. This allowed us to set
upper limits to the number density of KBOs that are stringent enough
to be compared 
usefully
with models for the formation and evolution of the
Kuiper Belt. We considered four theoretical models, PS05, KB04, BCB09
and F09, all of which describe the present size distribution of the
Kuiper Belt, and we set constraints on these models. This is the first
detailed comparison of occultation data with specific model results.

Our result, particularly when compared with PS05 and KB04, suggests
that the Kuiper Belt is populated by fragile bodies, and that the
effect of the migration of Neptune played an important role in its
formation. None of the BCB09 or F09 
models
can be ruled out. 


Using a 
generic model, where the size distribution is
described by three consecutive power laws, and fixing the slope on the
large end size and the location of the first break to the best fit
from direct observations ($q_\m{L}=4.8~\mkm,~D_\m{b1}=75~\mkm$), and
the slope of the small side to $q_\m{s}=4.0$ we can constrain
the intermediate slope $q_\m{I}$ and the location of the second break
$D_\m{b2}$, as shown in Figure~\ref{fig:3par_exc}. As direct surveys
are currently not sensitive to KBOs smaller than $D\sim~28~\mkm$
occultation surveys provide the only probe of this region of the size
spectrum, and the large TAOS dataset allowed us to set the first
constraints to the location of the second
break.

We also considered the Jupiter Family Comets. Assuming the JFCs are
injected into their present orbit from one of the Kuiper Belt
populations, Classical Belt, Plutinos, or from the Scattered Disk, we
compared the upper limit derived from our survey to the estimates of
the number of objects derived using the number of JFCs by
\citet{1997Icar..127...13L} for a population of cometary precursors
entirely in the CB, that of \citet{1997Icar..127....1M} for Plutinos
and of~\citet{2008ApJ...687..714V} for a progenitor population in the
SD. We can rule out the a unique precursor family composed of CB
objects $D\gtrsim3~\mkm$. This preliminary result confirms that
occultation surveys can help understanding the origin of JFCs.

A recent analysis of the HST guiding data (S09) reports the detection
of an occultation by a $D\sim1~\mkm$ KBO. This object is detected at
high ecliptic latiutde ($b\sim14\degree$). The surface density derived
in S09, assuming a straight power law size distribution for small
KBOs, is within the upper limit set by TAOS, although a $1\sigma$
increment over the best fit surface density is ruled out to better
than 95\% c.l.


TAOS has operated for over four years observing continuously with
three, and now four, 50~cm aperture telescopes. TAOS is only
marginally sensitive to sub-km KBO occultations, but we were able to
prove that the low sensitivity at sub-km sizes is more than
compensated for by the vast exposure of which TAOS is
capable. Improvements in our selection criteria and the analysis of
four telescope data should increase our sensitivity to smaller
objects.

\acknowledgements The authors wish to thank Scott Kenyon, for
insightful conversations. Work at the CfA was supported in part by the
NSF under grant AST-0501681 and by NASA under grant NNG04G113G.  Work
at NCU was supported by the grant NSC 96-2112-M-008-024-MY3. Work at
ASIAA was supported in part by the thematic research program
AS-88-TP-A02. Work at Yonsei was supported by National Research Foundation of Korea
through Grant 2009-0075376.
Space Science Institute. 
The work of N. Coehlo was supported in part
by NSF grant DMS-0636667.  Work at LLNL was performed in part under
USDOE Contract W-7405-Eng-48 and Contract DE-AC52-07NA27344. Work at
SLAC was performed under USDOE contract DE-AC02-76SF00515.  Work at
NASA Ames was supported by NASA's Planetary Geology \& Geophysics
Program.





\end{document}